\documentclass[aps,prd,preprint,floatfix,nobibnotes,nofootinbib]{revtex4-2}

\usepackage{amsmath}
\usepackage{amssymb}
\usepackage{graphicx}
\usepackage{float}
\usepackage{dsfont}
\usepackage{ulem}
\usepackage[colorlinks=true]{hyperref}
\usepackage{slashed}
\usepackage{MnSymbol}

\newcommand{\be}{\begin{equation}}
\newcommand{\ee}{\end{equation}}
\newcommand{\ba}{\begin{eqnarray}} 
\newcommand{\ea}{\end{eqnarray}} 
\newcommand{\nn}{\nonumber}

\newcommand{\bea}{\begin{eqnarray}}
\newcommand{\eea}{\end{eqnarray}}

\usepackage[usenames, dvipsnames]{color}

\numberwithin{equation}{section}

\usepackage{setspace}
\allowdisplaybreaks
\begin{document} 

\title{${\cal N}=4$ supersymmetric Yang-Mills thermodynamics to order $\lambda^{5/2}$}

\author{Margaret E. Carrington}
\affiliation{Department of Physics, Brandon University,
Brandon, Manitoba R7A 6A9, Canada}
\affiliation{Winnipeg Institute for Theoretical Physics, Winnipeg, Manitoba, Canada}

\author{Gabor Kunstatter}
\affiliation{Department of Physics, University of Winnipeg, Winnipeg, Manitoba, R3M 2E9 Canada}
\affiliation{Department of Physics, Simon Fraser University, Burnaby, British Columbia, V5A 1S6 Canada}
\affiliation{Winnipeg Institute for Theoretical Physics, Winnipeg, Manitoba, Canada}

\author{Ubaid Tantary} 
\affiliation{Department of Mathematics and Natural Sciences, Prince Mohammad Bin Fahd University, Al Khobar 31952, Saudi Arabia}

\date{\it April 07, 2026}

\begin{abstract}

We calculate the resummed perturbative free energy of ${\cal N} = 4$ supersymmetric
Yang-Mills in four spacetime dimensions (SYM$_{44}$) to order $\lambda^{5/2}$ in the 
't Hooft coupling at finite temperature and zero chemical potential. 
All infrared divergences cancel when we include contributions from SYM$_{44}$ ring diagrams and the final result is both ultraviolet and infrared finite. Our result has special significance since order $\lambda^{5/2}$ is the highest order calculation that can be done with perturbation theory, because there are nonperturbative effects associated with the magnetic mass scale that come into play at order $\lambda^3$. 
We compare results obtained with regularization by dimensional reduction (RDR), which preserves supersymmetry, and canonical dimensional regularization (DR). We also compare with a generalized Pad\'e approximant constructed by matching the weak coupling result at order $\lambda^2$ and the large $N_c$ strong coupling result at order $\lambda^{-3/2}$. Finally we make a comparison between our result and the QCD free energy and show that SYM$_{44}$ has better convergence properties. 
\end{abstract}

\maketitle

\newpage

%%%%%%%%%%%%%%%%%%%%%%%%%%%%%%%%%%%%%%%%%%%%%%%%%%%%%%%%
\section{Introduction}
\label{sec-intro}
%%%%%%%%%%%%%%%%%%%%%%%%%%%%%%%%%%%%%%%%%%%%%%%%%%%%%%%%

The ${\cal N} = 4$ supersymmetric theory in four spacetime dimensions (SYM$_{44}$) is the most famous example of a conformal field theory (CFT) in four dimensions. 
At finite temperature and in the weak coupling limit the theory has many similarities with quantum chromodynamics (QCD), but because of its high degree of symmetry it is also simpler to work with in some ways. 
For example, SYM$_{44}$ is ultraviolet finite at all orders in perturbation theory (the beta function is zero and the coupling constant does not run and is independent of the temperature). An important difference with QCD is that in the strong coupling limit the free energy can be calculated using the anti-de Sitter space (AdS)/CFT correspondence. 
For these reasons SYM$_{44}$ is often used as a model for hot QCD. 

In the strong coupling limit the ratio of the free energy to the ideal gas result is \cite{Gubser:1998nz}
\bea
\frac{\cal F}{{\cal F}_{\rm ideal}}= \frac{3}{4}\bigg[
1+\frac{15}{8}\zeta(3)\lambda^{-3/2} + {\cal O}(\lambda^{-3})
\bigg]\,\label{strong}
\eea
where $\lambda$ is the 't Hooft coupling. 
In the weak coupling limit the leading order contribution to the free energy is the ideal gas result and the perturbative corrections have the form
\bea
\frac{\cal F}{{\cal F}_{\rm ideal}} = 1 +a_2\lambda+a_3\lambda^{3/2}+\big(a_4+a'_4\log(\lambda)\big)\lambda^2 + a_5\lambda^{5/2} + {\cal O}(\lambda^3)\,.
\label{weak-expansion}
\eea
The subscripts on the coefficients in (\ref{weak-expansion}) correspond to the square of the associated power of $\lambda$. There is no $\lambda^{5/2}\log(\lambda)$ contribution and the reason is discussed in section \ref{results-sec}. 
The coefficients $\{a_2,a_3,a_4,a_4'\}$ are known \cite{Fotopoulos:1998es,Kim:1999sg,Vazquez-Mozo:1999yck,Nieto:1999kc,Blaizot:2000fc, 
Du:2020odw,Du:2021jai,Andersen:2021bgw}. In this paper we present our calculation of the coefficient $a_5$. 

The weak and strong coupling results should accurately give the free energy in their respective domains, but the radii of convergence of the two series is unclear and it is therefore difficult to know to what extent one or the other should be trusted at intermediate coupling. Various methods can be used to interpolate between the strong and weak coupling regimes, but again it is difficult to know how accurate these expressions are. One approach is to use a generalized Pad\'e approximant constructed so that the coefficients can be uniquely determined with information from the perturbative result to order $\lambda^2$ and the strong coupling result in (\ref{strong}). In section \ref{results-sec} we show that this function does not very accurately represent the weak coupling result to order $\lambda^{5/2}$. 

The coefficient $a_2$ in (\ref{weak-expansion}) can be computed in a straightforward way. Calculations at higher orders require a reorganization of perturbation theory. The method we use is called static resummation. It is a simplified version of the Braaten-Pisarski resummation that can only be used to calculate static quantities. 
The key is that static Green’s functions can be calculated directly in imaginary time and do not need to be analytically continued back to real time, which means
that we can use Euclidean propagators with discrete energies when analyzing infrared divergences. 

Before explaining further how the static resummation works we define the theory and our notation. 
All fields belong to the adjoint representation of the SU$(N_c)$ gauge
group. 
The gauge field is defined as in QCD. It can be expanded as $A_\mu=A_\mu^a t^a$ with colour generators $t^a$ that satisfy 
\bea
[t^a,t^b]=if_{abc}t^c ~~~\text{and}~~~ {\rm Tr}(t^a t^b)=\frac{1}{2}\delta^{ab}
\eea
where the indices $\{a,b,c,\dots\}\in(1,N_c^2-1)$ and the structure constants $f_{abc}$ are real and antisymmetric. We use the standard notation $N_c=C_A$ and $d_A=\delta_{aa}=N_c^2-1$. The covariant derivative is defined $D_\mu=\partial_\mu-ig[A_\mu,~]$ and the field strength tensor is $F_{\mu\nu} = \partial_\mu A_\nu - \partial_\nu A_\mu -ig [A_\mu,A_\nu]$. 
The 't Hooft coupling in eqs.~(\ref{strong}, \ref{weak-expansion}) is related to the coupling $g$ as  $\lambda=C_A g^2$. 
In the rest of this paper, except for the last section where we present our results, we use $g$ instead of $\lambda$. The reason is that for technical reasons the fractional exponents are more difficult to deal with in our calculation.
The fermionic fields are  four-component Majorana fermions. 
There are six real scalar fields written in a multiplet as $\Phi_A$ with $A\in(1,6)$. Three components are scalar and three are pseudoscalar so we can write $\Phi_A=(X_p,Y_q)$ with $\{p,q\}\in(1,3)$ where $X_p$ and $Y_q$ represent scalar and pseudoscalar fields, respectively. These fields can be expanded as $\Phi_A = \Phi_A^a t^a$, $X_p = X_p^a t^a$ and $Y_q = Y_q^a t^a$. 
The Lagrange density is:
\bea
{\cal L} = {\rm Tr}\,\bigg[\!&-&\frac{1}{2}F_{\mu\nu}F^{\mu\nu} 
+ (D_\mu\Phi_A)(D^\mu\Phi_A)+i\bar\psi_i \slashed{D}\psi_i \nn \\ &-&\frac{1}{2}g^2\left(i[\Phi_A,\Phi_B]\right)^2 
 -i g \bar\psi_i\left[\alpha^p_{ij}X_p + i \beta^q_{ij}\gamma_5Y_q,\psi_j\right]\,
\bigg] + {\cal L}_{\rm gf} + {\cal L}_{\rm gh}\,
\eea
where the six $4\times4$ matrices $\alpha^p$ and $\beta^q$ satisfy 
\bea
\{\alpha^p,\alpha^q\}=-2\delta^{pq}  ~~~\text{and}~~~
\{\beta^p,\beta^q\}=-2\delta^{pq}  ~~~\text{and}~~~
[\alpha^p,\beta^q]=0\,.
\eea
We work in covariant gauge with the Feynman choice for the gauge parameter. 
We calculate the integrands for all diagrams in Minkowski space and then rotate to Euclidean space to define the finite temperature sum-integrals. For technical reasons this is more convenient for us than starting with Euclidean space Feynman rules. The Minkowski space Feynman rules are given in appendix \ref{feyn-rules-sec}. 

The static resummation was introduced in \cite{Arnold:1992rz} and first used to calculate the free energy of pure gauge QCD at order $g^4$ in \cite{Arnold:1994ps}. It has since been used to obtain the free energy of full QCD at order $g^4$ in \cite{Arnold:1994eb}, QCD at order $g^5$ in \cite{Zhai:1995ac}, and SYM$_{44}$ at order $g^4$ in \cite{Du:2021jai}.
The main idea behind static resummation is that it provides a simple way to organize a perturbative calculation of integrals that depend on two scales. The temperature, $T$, is the hard scale. The hard thermal loop (HTL) longitudinal gluon mass, $m$, and scalar mass, $M$, are both of order $gT$ and  provide the soft scale. Momentum integrals are calculated by separating the hard and soft momentum regions. A propagator is hard if some components of its momentum are hard, and soft if all components are soft. A boson Matsubara frequency has the form $2n \pi T$ where $n$ is integer and therefore a boson propagator is soft only if the frequency is zero. Fermion propagators have frequencies $(2n+1)\pi T$ and are always hard because their frequencies cannot be zero.  
This means that the structure of the resummation should be determined by the behaviour of the boson propagators in the Euclidean infrared limit $(p_0=0, p\ll T)$ where their masses cannot be treated as perturbative corrections. We give the results for the HTL masses $m$ and $M$ in appendix \ref{int-results-sec}.
The Lagrange density is modified, in frequency space, as
\bea
{\cal L} = \left({\cal L}+ {\rm Tr}[m^2 A_0^a A_0^a \delta_{p_0} - M^2 \Phi^A\Phi^A \delta_{p_0}]\right)
-{\rm Tr}[m^2 A_0^a A_0^a \delta_{p_0} - M^2 \Phi^A\Phi^A \delta_{p_0}]\,
\label{static-resum}
\eea
where the notation $\delta_{p_0}$ indicates a Kronecker delta that will set the corresponding Matsubara mode to zero in Euclidean space.  The two terms in the round bracket are absorbed into the unperturbed Lagrangian and the last two terms are treated as perturbations and give counterterm vertices for gluons and scalars of the form
\bea
&&\text{gluons:~~~} -i\delta^{ab}g^{\mu 0}g^{\nu 0} m^2\delta_{p_0} \label{cterms}\\
&&\text{scalars:~~~~~} i\delta^{ab}\delta^{AB} M^2\delta_{p_0}\,.\nn
\eea
The modified unperturbed Lagrangian gives resummed propagators (in Feynman gauge) 
\bea
&& \text{gluon:~~}  
\Delta_{\mu \nu}^{ab}=-i \delta^{ab} \Big[\frac{g_{\mu \nu}}{P^2}+ \Big(\frac{1}{P^2-m^2}-\frac{1}{P^2}\Big)\delta_{p_0}g^{\mu 0}g^{\nu 0}\Big]  \label{resum-props} \\[4mm]
%A3 and A4 in 2105.02101
&& \text{scalar:~~}  \Delta^{ab}_{AB}= i \delta^{ab}\delta_{AB} \Big[\frac{1}{P^2}+\Big(\frac{1}{P^2- M^2}-\frac{1}{P^2}\Big)\delta_{p_0}\Big] \,.\nonumber
\eea
The fermion and ghost propagators and the nine non-counterterm vertices of the theory are given in appendix \ref{feyn-rules-sec}. 

For a one loop diagram the integration variable is denoted $P$, for two loop diagrams the  integration variables are $(P,K)$, and for three loop diagrams they are $(P,K,Q)$. 
We define $P^2=p_0^2+p^2$ and four dimensional dot products are sometimes written out as $P\cdot K = p_0k_0 + p\cdot k$ where $p\cdot k\equiv \vec p\cdot\vec k$. 
We calculate integrals in $d$  dimensions and use the conventional shorthand notation to denote summation over discrete frequencies and integration over three momenta:
\bea
 \text{bosons:~~~} \sumint_P &\equiv& \mu^{4-d} T \sum_{p_0=2\pi n T}\int\frac{d^{d-1}p}{(2\pi)^{d-1}} \label{matsu}\\
 \text{fermions:~~~} \sumint_{\{P\}} &\equiv& \mu^{4-d} T \sum_{p_0=\pi (2n+1) T}\int\frac{d^{d-1}p}{(2\pi)^{d-1}} \,\nn\\[4mm]
\int_p &\equiv& \mu^{4-d} \int\frac{d^{d-1}p}{(2\pi)^{d-1}}   \,.\nn 
\eea
We use modified minimal subtraction  which means that we use minimal subtraction with the renormalization scale $\bar\mu$ defined through $\mu^2=e^{\gamma_E}\bar\mu^2/(4\pi)$. 

We use the regularization by dimensional reduction (RDR) scheme \cite{Siegel:1979wq,Capper:1979ns}. One maintains supersymmetry by keeping fixed the size of the bosonic, fermionic, and scalar representations, denoted $D$, while the number of spatial dimensions is modified to regulate integrals. In our notation we fix $D=4$ and regulate momentum integrals  using $d=4-2\epsilon$. This prescription ensures that the cancellations between the bosonic and fermionic degrees of freedom necessary to maintain supersymmetry are automatically preserved. In section \ref{results-sec} we compare the results obtained with RDR and canonical dimensional regularization (DR). 

A simple example will illustrate the advantages of the static resummation method. We consider the contribution to the free energy from the scalar one loop diagram (the fourth diagram in fig.~\ref{1loop-fig}). If we had not used a static resummation, the Kronecker delta would be removed from equations (\ref{static-resum}, \ref{cterms}, \ref{resum-props}). The Euclidean scalar propagator would be $\delta_{ab}\delta_{AB}/(P^2+M^2)$ and the one loop contribution to the free energy would be \cite{jackiw}
\bea
%&& \frac{i}{P^2-M^2}\to \frac{1}{P^2+M^2} \nn \\
&& \Rightarrow -{\cal F}\sim -\frac{6d_A}{2}\sumint_P \log(P^2+M^2) 
= 6d_A\left(\frac{\pi^2 T^4}{90}-\frac{1}{24}T^2 M^2 + \frac{1}{12\pi} T M^3 +\dots\right)
\eea
where the dots represent higher order terms in the $M/T$ expansion. With the static resummation the Euclidean scalar propagator is 
\bea
&&D(P) = \delta_{ab}\delta_{AB} \left(\frac{1}{P^2}+\Big(\frac{1}{P^2+ M^2}-\frac{1}{P^2}\Big)\delta_{p_0}\right)  
= \delta_{ab}\delta_{AB} 
\left(
\frac{1-\delta_{p_0}}{P^2} + \frac{\delta_{p_0}}{p^2+M^2} 
\right)\,.\label{scalar-prop}
\eea
This propagator will sometimes be written
\bea
D(P)=\delta_{ab}\delta_{AB} \left(\frac{1}{P^2} + \delta_{p_0}\Delta_M(P)\right)\text{~~with~~}\Delta_M(P)\equiv \frac{1}{P^2+ M^2}-\frac{1}{P^2}
\label{delta-def}
\eea
and similarly we use $\Delta_m(P) \equiv (P^2+m^2)^{-1}-P^{-2}$. 

Using (\ref{scalar-prop}) we get that the contribution to the one loop free energy from the third diagram in fig.~\ref{1loop-fig} is
\bea
  -{\cal F}&\sim& -\frac{6d_A}{2} \left(\sumint_P (1-\delta_{p_0})\log(P^2) + T\int_p  \log(p^2+M^2) \right)\label{J0tilde} \\[2mm]
& =& 6d_A\left(\frac{\pi^2 T^4}{90} + \frac{1}{12\pi} T M^3 \right) \,.\nn
\eea
In the first integral the Kronecker delta can be thrown away since the resulting three dimensional integral has no scale and is zero in dimensional regularization. The second integral is straightforward to calculate, and clearly proportional to $M^3$ by dimensional analysis. The important point is that the static resummation cleanly separates the contributions from the hard and soft scales and gives a sum of two terms instead of a series expansion in $M/T$.

This example also illustrates an important general property of resummations in thermal field theory. The leading contribution to the free energy from any $n$-loop diagram is ${\cal O}(g^{2(n-1)})$ but there will be subleading terms, not necessarily analytic in $g^2$, that carry extra powers of $g$ (or $\log(g)$). In the example above, the leading contribution to the scalar one loop diagram is order $g^0$, but there are also subleading corrections (of order $g^3$ with a static resummation). 
This structure means that to calculate the free energy at order $g^4$ we need to include contributions to diagrams with $n\le 2$ loops from thermal masses, but the three loop diagrams can be calculated with hard bare propagators because they carry an explicit factor $g^4$. At order $g^5$ we do not need four loop diagrams (because they have an explicit $g^6$), but we do need subleading contributions to the three loop diagrams. 
This is in fact the highest order that can be calculated in perturbation theory. At four loops there are infrared problems associated with the magnetic scale and a nonperturbative ${\cal O}(g^6)$ contribution to the free energy \cite{joe-charles,Nieto:1999kc}. Our calculation is the best that can ever be done with perturbation theory.

\section{Method to simplify amplitudes}
\label{method-sec}

\subsection{Organization}
\label{organiz-sec}

The full set of diagrams that we need to calculate are shown in figs.~\ref{1loop-fig}-\ref{gl-sc-base-fig}. The one loop graphs and one loop counterterms are shown in fig~\ref{1loop-fig}, fig~\ref{2loop-fig} is the two loop graphs, fig.~\ref{3loop-fig} shows the three loop graphs not including gluon and scalar baseballs, and fig.~\ref{gl-sc-base-fig} shows the gluon and scalar baseballs together with their counterterm contributions. 

We use a well established method to calculate all of these amplitudes. The calculation is difficult because there are so many diagrams. This not only makes the calculation very lengthy, it also means that it is very difficult to avoid errors, and that there is no simple way to identify and fix any possible mistakes. To circumvent these problems we have written a Mathematica program to perform all steps of the calculation. The basic idea is to use variable transformations to manipulate all terms in a given amplitude into simpler forms. These simplified expressions can then be matched with finite temperature sum-integrals that have been calculated previously in the literature (we will need to introduce only three new integrals). The program correctly reproduces the order $g^4$ SYM$_{44}$ result of \cite{Andersen:2021bgw,Du:2021jai} and the order $g^5$ QCD result of \cite{Zhai:1995ac}. Our order $g^5$ SYM$_{44}$ calculation is more complicated than either of these previous calculations, but the fact that our result is infrared and ultraviolet finite is further strong evidence of the reliability of our method. There are several other checks that are listed in section \ref{results-sec}. We emphasize that while no part of our method is original, what we have done is to invent a procedure to efficiently apply a series of tricks invented by other authors to any diagram up to three loop order. The entire process can take anywhere from 10 minutes up to several hours for a given diagram but the procedure itself is reasonably straightforward. The basic structure of the method is described below and further details are given in the appendices. 
There are two resummed propagators (see eq.~(\ref{resum-props})) that contain the factors $\Delta_m$ and $\Delta_M$ and these factors play a crucial role in the organization of the calculation. For purposes of discussion we sometimes consider only $\Delta_m$ and suppress the subscript $m$. The main steps in our process are as follows.

\begin{enumerate}
\item
For each diagram, construct the amplitude using the Feynman rules and contract all indices. This is a tedious process but can be done with a symbolic computation program, like FeynCalc or FORM. The amplitude for a three loop diagram has typically several hundred terms but the largest has over 4,000 terms.  After the amplitude has been calculated we rotate to Euclidean space. Symmetry factors are calculated using the method in ref.~\cite{Palmer:2001vq}. 

\item Perform variable transformations so that all Kronecker deltas depend only on the integration variables. For example, if a term has the product $\delta_{p_0}\delta_{p_0+k_0}$ then a transformation $K\to-K-P$ can be used to rewrite it so that the Kronecker deltas have the form $\delta_{p_0}\delta_{k_0}$. 

\item For each amplitude, but excepting the counterterm contributions in fig.~\ref{gl-sc-base-fig}, divide all terms into groups according to how many Kronecker deltas there are: a two loop diagram has terms of type$_0$, type$_1$ and type$_2$ (corresponding to 0 or 1 or 2 Kronecker deltas), and a three loop diagram has terms of type 0, 1, 2, or 3. 

\item The counterterm contributions in fig.~\ref{gl-sc-base-fig} are treated slightly differently because for these terms there is an `extra' Kronecker delta coming from the counterterm itself (see eq.~(\ref{cterms})) instead of from the resummed propagator. These Kronecker deltas do not bring an accompanying factor of $\Delta$. 
Counterterm contributions are grouped into type$_{10}$, type$_{11}$, type$_{21}$ and type$_{22}$ where the first index gives the number of Kronecker deltas and the second is the number of factors of $\Delta$. 

\item Rewrite each term in the integrand in terms of the integrals given in appendix~\ref{int-def-sec}. 
This is the non-trivial step and we give further details about how it is done in the following subsections and in the appendices. 
In general it requires a series of variable transformations and symmetrization operations.
For some of the three loop integrals we must also perform an expansion to extract the leading and next-to-leading pieces of the integral. 

\end{enumerate}

\subsection{One and two loop diagrams}
\label{two-loop-sec}
All of the terms in each one loop and two loop diagram can be written in the form of the integrals in equations (\ref{1-hard}, \ref{2-hard}, \ref{2-soft-hard}, \ref{1-soft}, \ref{2-soft}) by shifting integration variables. The integrals in (\ref{2-soft-hard}) are written in terms of the others using the identities in section \ref{identities-sec} which were derived in \cite{Zhai:1995ac}. The results for the remaining integrals are given in section \ref{int-results-sec} and are taken from \cite{Arnold:1994ps,Arnold:1994eb}. 

\subsection{Non-counterterm three loop contributions}
\label{dividing-sec}

The motivation for separating these terms according to the number of Kronecker deltas they have is as follows. 
From the form of the resummed propagators in eq.~(\ref{resum-props}) we know that each Kronecker delta comes with an associated factor $\Delta$. In a three loop diagram, a factor $\delta_{p_0}\Delta(P)$ forces the momentum $p$ to be soft. The reason is that if $p$ were not soft, the three loop amplitude would be suppressed by a factor $m^2/p^2$ and would give a contribution of order ${\cal O}(g^4 \times g^2)={\cal O}({g^6})$. 
From this argument one might guess that counting the number of Kronecker deltas is equivalent to counting the number of factors $\Delta$, and that the number of $\Delta$'s is the number of soft variables. Neither of these is always correct.  

First consider the possible factors
\bea
&& [\delta_{k_0} \Delta(K)]^2=\delta_{k_0} [\Delta(K)]^2 \label{ex1} \\ 
&& \delta_{p_0}\delta_{k_0}\delta_{p_0+k_0} \Delta(P)\Delta(K)\Delta(P+K) = \delta_{p_0}\delta_{k_0} \Delta(P)\Delta(K)\Delta(P+K)  \label{ex2} \\ 
&& \delta_{p_0}\delta_{k_0} \delta_{q_0} \delta_{p_0+k_0+q_0}\Delta(P)\Delta(K)\Delta(Q)\Delta(P+K+Q) \label{ex3} 
 =\delta_{p_0}\delta_{k_0} \delta_{q_0} \Delta(P)\Delta(K)\Delta(Q)\Delta(P+K+Q)\,. \nn
\eea
These examples show that the number of Kronecker deltas is less than or equal to the number of factors of $\Delta$. 
Next we note that while each Kronecker delta gives a soft variable, the number of soft variables can be greater than the number of Kronecker deltas. This happens because any variables that do not enter through a factor $\delta_{x_0}\Delta(X)$ (for example, $P$ and $Q$ in  eq.~(\ref{ex1}) and $Q$ in eq.~(\ref{ex2})) could give contributions from both hard and soft regions. 

Now we explain how to calculate terms of each type, where `calculate' means to rewrite in the form of one of the known integrals in appendix \ref{int-def-sec}. The first step in the process is to `decouple' the integrals, which means to separate contributions from hard and soft regions. The process is described in general terms below and in appendix \ref{separation-sec} we give a example of decoupling a type$_1$ term. 

For type$_0$ terms all momenta are hard so they are already decoupled. There is an explicit factor $g^4$ from the vertices of the diagram so all of these terms contribute at order $g^4$. 

For type$_3$ terms all variables are soft, since all frequencies are set to zero and the only remaining scale is the soft scale. These integrals are therefore also already decoupled. From dimensional analysis it is easy to see that they contribute at order $g^4  g^{9} g^{-8} = g^5$ where $g^4$ is the explicit factor from the vertices of the diagram, $g^9$ comes from three 3-dimensional momentum integrals, and $g^{-8}$ gives the integrand its correct dimensions since we have $12$ factors with dimension $T$ from the three sum-integrals and we must end up with overall dimension four. 

Terms that are type$_2$ have a factor of the form $\delta_{p_0}\delta_{k_0} \Delta(P)\Delta(K)$. If the third variable, $Q$, is coupled to either $P$ or $K$ we must decouple it. To do this we rewrite the product of Kronecker deltas as 
\bea
\delta_{p_0}\delta_{k_0} = \delta_{p_0}\delta_{k_0}[1-\delta_{q_0}]+\delta_{p_0}\delta_{k_0} \delta_{q_0}\label{type2-split}\,.
\eea
Everything multiplying the second term on the right with the three Kronecker deltas is extracted and regrouped with the original type$_3$ terms. The dominant contribution from the first piece on the right side of (\ref{type2-split}) comes from $(p,k)$ soft and $Q$ hard, which means we can expand in $(p,k)\ll Q$. This expansion decouples the integrals. These terms are called truetype$_2$ because they have 2 soft integration variables (instead of two Kronecker deltas and two or three soft integration variables). 

Similarly we rewrite type$_1$ terms using 
\bea
\delta_{p_0} = \delta_{p_0}[1-\delta_{k_0}\delta_{q_0}]+\delta_{p_0}\delta_{k_0} \delta_{q_0}\label{type1-split}\,.
\eea
Everything multiplying the second term on the right is extracted and added to other type$_3$ terms. The dominant contribution to the first piece on the right side of (\ref{type1-split}) comes from $p$ soft and $(K,Q)$ hard. We can therefore expand in $p\ll (K,Q)$ and this decouples the integrals. These terms are called truetype$_1$ because they have one soft integration variable. 

\subsection{Counterterm three loop contributions}

These terms have to be handled slightly differently. 
The reason is that there are two ways Kronecker deltas can appear in counterterm contributions. 
One is from a resummed propagator, exactly as for non-counterm contributions. 
In addition there are also deltas coming from the counterterm insertion itself (see eq.~(\ref{cterms})). 
As explained in section \ref{organiz-sec}, we group these terms according to the number of Kronecker deltas and the number of factors of $\Delta$.

Terms of type$_{10}$ have a Kronecker delta from a counterterm insertion but no $\Delta$ factors. For these integrals all momenta are hard and they do not need to be decoupled. %cannot expand, ct or ct^2

Terms of type$_{11}$ have a factor $\delta_{p_0}\Delta(P)$ (or $\delta_{p_0}\Delta^2(P)$). At least one of the remaining momenta will be uncoupled from the other two, from the construction of the counterterm. If we call this momentum $Q$ then the remaining momentum is $K$, and since $K$ is not soft it can be decoupled from the soft $p$ by expanding in $p/K$. %must expand, ct or ct^2

Terms of type$_{21}$ have a factor $\delta_{p_0}\delta_{k_0}\Delta(P)$. The third momentum variable, $Q$, is decoupled as before. Since the $\delta_{k_0}$ does not come from a resummed propagator we expand in $p/K$. %must expand, ct

Terms of type$_{22}$ have a factor $\delta_{p_0}\delta_{k_0}\Delta(P)\Delta(K)$ and are already decoupled.  %expanding is irrelevant, ct

\subsection{Identifying fundamental integrals}

After all terms have been written in terms of decoupled hard and soft integrals we can rewrite each of them in terms of the fundamental integrals in section \ref{int-def-sec}. 
The basic strategy is to apply variable transformations to shift and/or rename the integration variables until the term matches one (or sometimes more than one) of the known fundamental integrals. The part of the program that does this receives as input the set of all transformations that are allowed for a given term. There is a different set of transformations according to what variables are bosonic or fermionic. These assignments are made when the amplitude for the diagram is originally calculated (for example, using FeynCalc) and must be `remembered' when further processing is done. In practice, a single index is assigned to each case, for example the index 1 means we have a two loop diagram with bosonic variables $(P,K)$, 2 means we have a two loop diagram with $P$ bosonic and $K$ fermionic, 3 means $(P,K,Q)$ are bosonic, 4 means $(P,K)$ are bosonic and $Q$ is fermionic, etc. 
This index is passed to all further processing modules so that the correct symmetries are used. 

The first step is to expand all factors of $\Delta$. They are no longer needed once we have separated each term into hard and soft integration regions and the fundamental integrals in appendix \ref{int-def-sec} are written in terms of propagators only. Some of the terms this expansion produces can be dropped because they have no scale and are set to zero in dimensional regularization, but the number of terms in a given amplitude is still much greater after expanding the $\Delta$'s.

After the $\Delta$ factors have been expanded the program searches through the set of allowed variable transformations it is given, looking for a form that matches one of the fundamental integrals listed in appendix \ref{int-def-sec}. For some terms the match is easy to find, two examples are given in equations (\ref{sh1}, \ref{sh2}). In other cases one needs both variable transformations and symmetrization operations to find the desired form. 
In practice the key is to remove dot products from numerators whenever possible. There are only four fundamental integrals with dot products in their numerators ($H_3$, $H_4$, $H_5$, $H_6$) and all other dot products must be removed. Several different strategies are employed to remove all the unwanted dot products from the numerators. These different methods are explained in appendix \ref{reduction-sec} with examples for each. 
After all dot products have been removed from numerators, the program tries all of the allowed transformations until it finds the form of one of the fundamental integrals. 

Results for each individual diagram are given in appendix \ref{results-diags-sec}. 

\section{Results}
\label{results-sec}
In the weak coupling limit the ratio of the free energy to the ideal gas free energy can be written as  in eq.~(\ref{weak-expansion}). To evaluate the coefficients we sum the contributions from the individual diagrams in section \ref{results-diags-sec} and substitute the results for each of the integrals from section \ref{int-results-sec}. Our final result written in terms of the 't Hooft coupling $\lambda=C_A g^2$ is
\bea
\frac{\cal F}{{\cal F}_{\rm ideal}}&=& 1 +a_2\lambda+a_3\lambda^{3/2}+(a_4+a'_4\log(\lambda))\lambda^2
+a_5\lambda^{5/2}
\label{weak-expansion-2}  \\
&& \hspace*{-1cm} =  1 -\frac{3}{2 \pi ^2} \lambda 
+ \left(3+\sqrt{2}\right) \frac{\lambda^{3/2} }{\pi^3} \nn \\[2mm]
&& \hspace*{-1cm} + \left(
\frac{3 \zeta '(-1)}{2 \zeta (-1)}+\frac{3 \log (\lambda)}{2}+\frac{3 \gamma }{2}-\frac{9}{4
   \sqrt{2}}-\frac{21}{8}-3 \log (\pi )-\frac{25 \log (2)}{8}
\right)\frac{\lambda^2}{\pi^4} \nn \\[2mm]
%
%&& \hspace*{-1cm} + \left(\frac{1197}{256}
%   (\log (4)-1)+\frac{9}{256} \left(\sqrt{2}-44\right) \log \left(1+\sqrt{2}\right)-\frac{1}{64} \left(6+\sqrt{2}\right) \pi ^2-\frac{44+53 \log (2)}{32 \sqrt{2}}\right)\frac{\lambda^{5/2}}{\pi^5}\,. \nn \\
&& \hspace*{-1cm} + \left(
\frac{33}{8} (\log (4)-1)
+\frac{3}{128} \left(28+25
   \sqrt{2}\right) \log \left(1+\sqrt{2}\right)
-\frac{1}{64} \left(6+\sqrt{2}\right) \pi ^2-\frac{35+212 \log (2)}{128
   \sqrt{2}}
\right)\frac{\lambda^{5/2}}{\pi^5}\,. \nn 
\eea

Before discussing our result in more detail we explain some of the checks we have performed. 
We recognize that when a lengthy and complicated calculation is done using Mathematica it is natural to worry that there could be a mistake. We list below a number of reasons that we think our calculation is correct. 

\begin{enumerate}
\item The result is finite. This happens because of explicit cancellations between three loop infrared singularities and the three loop counterterm diagrams. There are no poles from ultraviolet divergences because the coupling does not run in SYM$_{44}$ and therefore no coupling constant renormalization
counterterm is needed. 

\item Our result to order ${\cal O}(\lambda^2)$ agrees with the result that was calculated previously in \cite{Du:2021jai,Andersen:2021bgw}. Both papers use different methods. The first uses a static resummation but works in a 10 dimensional space with a reduced set of diagrams, and the second uses an effective field theory method. The important point is that in our calculation 
the set of diagrams that contribute at order $\lambda^2$ and $\lambda^{5/2}$ is exactly the same 
(the difference is that at order $\lambda^2$ the three loop amplitudes can all be calculated with bare propagators). The fact that we reproduce the previous order $\lambda^2$ result is evidence that we have the correct overall factor for each diagram. 

\item We have used our program to calculate the QCD free energy at order $g^5$ and our result agrees with the previous result of ref.~\cite{Zhai:1995ac}. The QCD calculation involves a much smaller set of diagrams but the fundamental  integrals that are needed are the same as in our calculation, except that we need three additional integrals $(J_{\rm 3n}, J_{\rm 3m}, J_{\rm 3p})$, that reduce to one of the QCD forms when $M=m$. The fact that we reproduce the previous QCD order $g^5$ result is evidence that our process correctly divides three loop amplitudes into hard and soft regions and matches each term to a known integral at this order. 

\item In the final result, for all terms that involve three loop integrals, if one momentum integral is decoupled from the other two, then the remaining two are always decoupled from each other. This is called the cancellation of double overlapping sum-integrals. The behaviour was seen in the QED free energy in \cite{Parwani:1994xt} and also occurs in the order $g^5$ QCD result of \cite{Zhai:1995ac}. 

\item The absence of terms $\sim g^{2n+1}\log(g)$ is expected (in our calculation the absence of a $g^5\log(g)$ contribution). This can be understood within the effective field theory picture as follows \cite{Braaten:1995cm}. 
Imagine first integrating out the nonstatic fields (scale $T$ physics) to arrive at an effective field theory which correctly describes physics in the low energy region (of order $gT$). Let $\Lambda$ be the cutoff separating the scales $T$ and $gT$. All contributions to the free energy from integrating out the hard modes have even powers of $g$ since there is no resummation. A $g^{2n}\log(g)$ term can arise 
through a cancellation between $\log(\Lambda/T)$ and $\log(\Lambda/(gT)$ terms.
Integrating out the nonstatic fields generates effective interaction terms for the static
fields. In principle this could introduce a dependence on the cutoff $\Lambda$ into the bare parameters of the effective field theory for the static fields at scale $gT$. A $g^{2n}g \log(g)$ term could then arise through a cancellation between a factor $\log(\Lambda/T)$ in an effective parameter and $\log(\Lambda/(gT)$ terms in a calculation from the effective theory.  Since neither the mass nor the coupling is renormalized in our theory these terms cannot appear. In fact the same holds in QCD \cite{Zhai:1995ac}. 

\end{enumerate}

Our calculation is done using the RDR scheme which manifestly preserves supersymmetry. It is of interest to see how the result would change using canonical DR. This is simple for us to check since it only involves changing one parameter in our program. If we use $D=d=4-2\epsilon$ instead of $D=4$ the result in (\ref{weak-expansion-2}) becomes
\bea
\frac{\cal F}{{\cal F}_{\rm ideal}}&=&  1 +a_2\lambda+a_3\lambda^{3/2}+(a_4+a'_4\log(\lambda))\lambda^2
+a_5\lambda^{5/2}
\label{weak-expansion-3} \\
&& \hspace*{-1cm}  =  1 -\frac{3}{2 \pi ^2} \lambda 
+ \left(3+\sqrt{2}\right) \frac{\lambda^{3/2} }{\pi^3} \nn \\
&& \hspace*{-1cm}  + \left(
\frac{3 \zeta '(-1)}{2 \zeta (-1)}
+\frac{3 \log (\lambda)}{2 }+\frac{3 \gamma }{2}-\frac{9}{4 \sqrt{2}}
-\textcolor{blue}{\frac{369}{128}}-3 \log (\pi )
-\frac{25
   \log (2)}{8}
\right)\frac{\lambda^2}{\pi^4} \nn \\
%
%&& \hspace*{-1cm}  + \left(
% \frac{1197}{256}(\log (4)-\textcolor{blue}{\frac{117}{133}})
%+\frac{9}{256} \left(\sqrt{2}-44\right) \log \left(1+\sqrt{2}\right)
%-\frac{1}{64} \left(6+\sqrt{2}\right) \pi ^2-\frac{\textcolor{blue}{38}+53 \log (2)}{32 \sqrt{2}}
%\right)\frac{\lambda^{5/2}}{\pi^5}\,. \nn \\
&& \hspace*{-1cm} + \left(
\frac{33}{8} (\log (4)-\textcolor{blue}{\frac{19}{22}})
+\frac{3}{128} \left(28+25
   \sqrt{2}\right) \log \left(1+\sqrt{2}\right)
-\frac{1}{64} \left(6+\sqrt{2}\right) \pi ^2-\frac{\textcolor{blue}{11}+212 \log (2)}{128
   \sqrt{2}}
\right)\frac{\lambda^{5/2}}{\pi^5}\,. \nn 
\eea
There are three changed coefficients which are marked in blue to make them easier to see. 
%The effect of these changes is too small to be seen on a plot at order $\lambda^{5/2}$. 

\begin{figure}[H]
\begin{centering}
\includegraphics[scale=1.3]{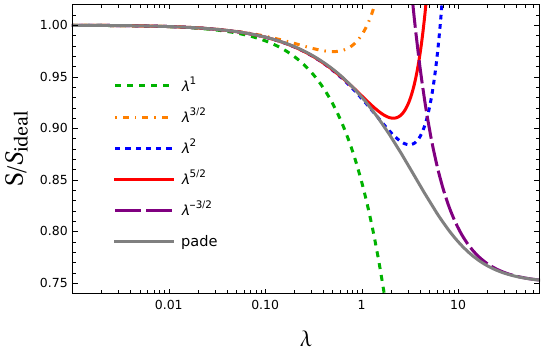}
\caption{The scaled entropy as a function of coupling at different orders in perturbation theory are shown in green (dashed), orange (dot-dashed), blue (dotted) and red (solid). The purple (long-dashed) line is the result at order $\lambda^{-3/2}$. The gray line joining the blue and purple lines is the generalized Pad\'e approximant in eq.~(\ref{pade}). \label{entropy-fig}}
\end{centering}
\end{figure}

We also discuss the use of a generalized Pad\'e approximant to interpolate between the weak and strong coupling results. This approximant was constructed in \cite{Blaizot:2000fc} using the information available from the weak coupling expansion of the free energy to order $\lambda^{3/2}$. In ref.~\cite{Du:2021jai} it was extended  using the new constraints available from their calculation of the perturbative free energy at order $\lambda^2$. One defines
\bea
\frac{{\cal F}}{{\cal F}_{\rm ideal}} = \frac{1+a\lambda^{1/2}+b\lambda+c\lambda^{3/2}+d \lambda^2+e\lambda^{5/2}}{1+\tilde a\lambda^{1/2}+\tilde b\lambda+\tilde c\lambda^{3/2}+\tilde d \lambda^2+\tilde e\lambda^{5/2}}\,.\label{pade}
\eea
The 10 coefficients in this expression are uniquely determined from 10 conditions. Expanding (\ref{pade}) in $\lambda$ and matching to the known coefficients of the terms of order $(\lambda^{1/2},\lambda,\lambda^{3/2},\lambda^2)$ gives four equations. Then one expands (\ref{pade}) in $1/\lambda$ and matches to the strong coupling result (\ref{strong}). In the large $N_c$ limit the strong coupling expansion becomes a series in powers of $\lambda^{-3/2}$. One therefore has six equations from the coefficients of the terms of order $(\lambda^0,\lambda^{-1/2},\lambda^{-1},\lambda^{-3/2},\lambda^{-2},\lambda^{-5/2})$.  The results for the coefficients of the function in (\ref{pade}) are given in appendix G of \cite{Du:2021jai} and the curve is plotted in fig.~\ref{entropy-fig}. Note that the approximant in (\ref{pade}) cannot be updated with information from our order $\lambda^{5/2}$ result because we would need two conditions to determine two additional coefficients from terms $f\lambda^3$ and $\tilde f\lambda^3$. 

The pressure, entropy density, and energy density can be obtained from the free energy using 
standard thermodynamic identities:
${\cal P} = -{\cal F}$, ${\cal S} = -d{\cal F}/dT$ and  ${\cal E} = {\cal F} - T d{\cal F}/dT.$ For SYM$_{44}$ the ratios of all of these quantities with the corresponding ideal gas result are the same. This happens because the free energy  depends on $T$ only through the prefactor ${\cal F}_{\rm ideal}\sim T^4$ (due to the conformality of SYM). 
In fig.~\ref{entropy-fig} we plot the ratio ${\cal F}/{\cal F}_{\rm ideal}={\cal S}/{\cal S}_{\rm ideal}={\cal P}/{\cal P}_{\rm ideal}={\cal E}/{\cal E}_{\rm ideal}$ as a function of $\lambda$. The perturbative results at different orders in $\lambda$ are shown in green (dashed), orange (dot-dashed), blue (dotted) and red. The purple curve shows the order $\lambda^{-3/2}$ strong coupling result. The gray line is the Pad\'e approximant in eq.~(\ref{pade}). The graph shows that while the Pad\'e approximant interpolates smoothly between the order $\lambda^2$ weak coupling result and the strong coupling expression, as it was designed to, it does not match well with the weak coupling result to order $\lambda^{5/2}$. 

\begin{figure}[H]
\begin{centering}
\includegraphics[scale=0.8]{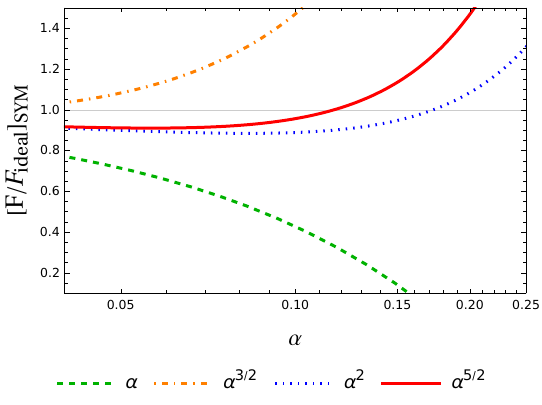}~~
\includegraphics[scale=0.8]{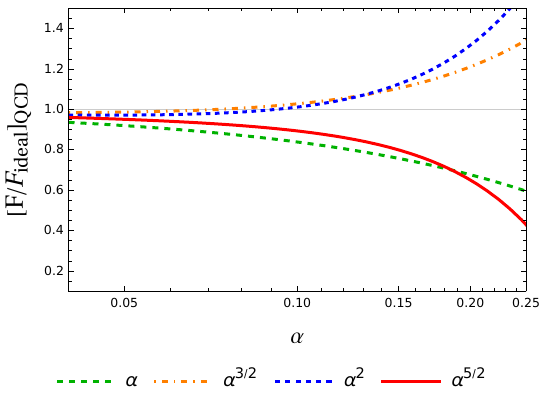}
\caption{The SYM$_{44}$ free energy as a function of $\alpha=\lambda/(4\pi C_a)$ and the QCD free energy with $\bar\mu = 2\pi T$ as a function of $\alpha=g^2/(4\pi)$ with $N_f=N_c=3$.\label{QCD-fig}}
\end{centering}
\end{figure}

To compare the convergence properties of SYM$_{44}$ and QCD we show in fig.~\ref{QCD-fig} the free energies of both theories as a function of the coupling $\alpha=g^2/(4\pi)$. The figure shows that the convergence of SYM$_{44}$ is better than that of QCD, which might be related to the higher degree of symmetry in the supersymmetric theory. 

\begin{figure}[H]
\begin{centering}
\includegraphics[scale=1.2]{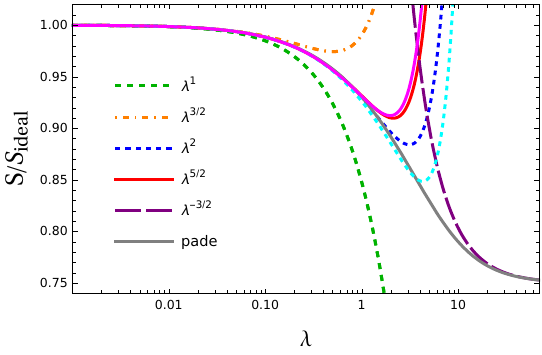}
\caption{The SYM$_{44}$ free energy as shown in fig.~\ref{entropy-fig} and the fourth order (cyan) and fifth order (magenta) results obtained with DR. \label{DR-fig}}
\end{centering}
\end{figure}

In fig.~\ref{DR-fig} we show our result for the free energy in eq.~(\ref{weak-expansion-2}) and the result obtained with DR in eq.~(\ref{weak-expansion-3}). The figure shows that the difference at order $\lambda^{5/2}$ is small for the range of couplings we consider. We remind the reader that only the RDR scheme preserves supersymmetry and the result obtained with DR is only of interest because it provides a check of the sensitivity of the calculation to the regularization. 
\section{Conclusions}
\label{conclusions-sec}

In this paper we have computed the thermodynamic function(s) of SYM$_{44}$ at order $\lambda^{5/2}$. 
The result extends our knowledge of weak coupling SYM$_{44}$ thermodynamics. 
Our order $\lambda^{5/2}$ result is particularly important because it clearly shows that the perturbative convergence of the free energy is better for SYM$_{44}$ than for QCD. This property is likely due to the larger degree of symmetry in the supersymmetric theory but is not conclusively demonstrated at lower orders. Our calculation also has special significance because it is the highest order perturbative calculation that can be done before nonperturbative effects related to magnetic scales become relevant \cite{Nieto:1999kc}. 

\appendix

\section{Feynman rules}
\label{feyn-rules-sec}

There are four propagators
\bea
% A31 in 2006.02617
1. && \text{fermion propagator} = \frac{i \delta^{ab}\delta_{ij}\slashed{P}}{P^2} \nonumber \\
2. && \text{ghost propagator} = i\frac{\delta^{ab}}{P^2} \nonumber \\
%
%A1 and A2 in 2105.02101
3. && \text{gluon propagator (Feynman gauge)} = 
-i \delta^{ab} \Big[\frac{g_{\mu \nu}}{P^2}+ \Big(\frac{1}{P^2-m^2}-\frac{1}{P^2}\Big)\delta_{p_0}g^{\mu 0}g^{\nu 0}\Big]  \nonumber \\
%A3 and A4 in 2105.02101
4. && \text{scalar propagator} = i \delta^{ab}\delta^{AB} \Big[\frac{1}{P^2}+\Big(\frac{1}{P^2- M^2}-\frac{1}{P^2}\Big)\delta_{p_0}\Big] \nonumber
\eea

There are nine vertices (all momenta are ingoing)
\bea
1. && \text{quark gluon vertex (gluon/fermion-out/fermion-in)} \nonumber \\
&& \Gamma^{\mu ij}_{abc}(p,k,q)= g \gamma^\mu f_{abc} \delta_{ij} \nonumber \\
2. && \text{gluon 3-vertex} \nonumber \\
&& \Gamma_{abc}^{\mu\nu\rho}(P,K,Q) =   
 g f_{abc}[ g^{\mu\nu}(P-K)^\rho + g^{\nu\rho}(K-Q)^\mu+g^{\rho\mu}(Q-P)^\nu ] \nonumber \\
3. && \text{gluon 4-vertex} \nonumber \\
&&  \Gamma_{abcd}^{\mu\nu\rho\sigma}(P,K,Q,R) =    \nonumber \\
&& - i g^2 [ f_{abe}f_{cde}(g^{\mu\rho}g^{\nu\sigma} - g^{\mu\sigma}g^{\nu\rho})  + f_{ace}f_{bde}(g^{\mu\nu}g^{\rho\sigma} - g^{\mu\sigma}g^{\nu\rho}) + f_{ade}f_{bce}(g^{\mu\nu}g^{\rho\sigma} - g^{\mu\rho}g^{\nu\sigma})]\nonumber \\
4. && \text{ghost gluon vertex (gluon/ghost-out/ghost-in)} \nonumber \\
&& \Gamma_{abc}^{\mu}(P,K,Q) = - g f_{abc} K^\mu \nonumber \\
5. &&\text{scalar 4-vertex} \nonumber \\
&& \Gamma^{ABCD}_{abcd}(p,k,q,r) = \nonumber \\
&& \tiny{-i g^2 [ f_{abe}f_{cde}(\delta^{AC}\delta^{BD} - \delta^{AD}\delta^{BC})  + f_{ace}f_{bde}(\delta^{AB}\delta^{CD} - \delta^{AD}\delta^{BC}) + f_{ade}f_{bce}(\delta^{AB}\delta^{CD} - \delta^{AC}\delta^{BD})]} \nonumber \\
6. && \text{1 gluon - 2 scalars}\nonumber \\
&& \Gamma^{\mu A B}_{abc}(P,K,Q) =  -g f_{abc} \delta^{AB} (K-Q)^\mu \nonumber \\
7. && \text{2 gluon - 2 scalar} \nonumber \\
&& \Gamma^{\mu\nu AB}_{abcd}(P,K,Q,R) = i g^2 g^{\mu\nu} \delta^{AB} (f_{ade} f_{bce} + f_{ace} f_{bde})  \nonumber \\
8. && \text{scalar fermion (scalar/fermion-out/fermion-in)}\nonumber \\
&&   \Gamma^{pij}_{abc}(P,K,Q) =
-i g f_{abc} \alpha_{ij}^p \nonumber \\
9. && \text{pseudo-scalar fermion (pseudo-scalar/fermion-out/fermion-in)} \nonumber \\
&&  \Gamma^{qij}_{abc}(P,K,Q) =
 g f_{abc} \beta_{ij}^q \gamma_5 \,.\nonumber
\eea

\section{Separation example}
\label{separation-sec}

As explained in section \ref{method-sec}, three loop type$_1$ and type$_2$ terms are rewritten as decoupled hard and soft integrals, neglecting terms order $g^6$. 
In this section we give an example of how to decouple a type$_1$ term.

A typical type$_1$ term is shown in eq.~(\ref{example-type1})
\bea
g^4\sumint_{PKQ}\frac{\delta_{p_0}}{K^2 Q^2 (P+K+Q)^2(P^2+m^2)}\,. \label{example-type1}
\eea
We divide the term into two pieces by rewriting it in the form
\bea
g^4\sumint_{PKQ}\frac{\delta_{p_0}\delta_{k_0}\delta_{q_0}}{K^2 Q^2 (P+K+Q)^2(P^2+m^2)}+g^4\sumint_{PKQ}\frac{\delta_{p_0}(1-\delta_{k_0}\delta_{q_0})}{K^2 Q^2 (P+K+Q)^2(P^2+m^2)}\,. \label{example-type1b}
\eea
The first term in (\ref{example-type1b}) is now type$_3$ so it is removed and regrouped with the other type$_3$ terms. The second term is dominated by $(K,Q)$ hard and therefore can be approximated as  
\bea
&& g^4 \sumint_{PKQ}\frac{\delta_{p_0}(1-\delta_{k_0}\delta_{q_0})}{K^2 Q^2 (K+Q)^2(P^2+m^2)} +{\cal O}(g^6)\nn \\[2mm]
&=&g^4\sumint_{KQ}\frac{(1-\delta_{k_0}\delta_{q_0})}{K^2 Q^2 (K+Q)^2}\sumint_P\frac{\delta_{p_0}}{(P^2+m^2)} +{\cal O}(g^6) \nn \\[4mm]
&=&  g^4 I_{\rm bsun} \, J_{\rm m1a}  +{\cal O}(g^6)
\eea
where the term proportional to $\delta_{k_0}\delta_{q_0}$ in the second line is dropped because it is zero in dimensional regularization. 

%%%%%%%%%%%%%%%%%%%%%%%%%%%%%%%%%%%%%%%%%%%%%%%%%%%%%%%%%%%%%%
\section{Reduction tricks}
\label{reduction-sec}

In most of the  examples we give in this section  we take all variables to be  bosonic because the most complicated cases occur when there are three bosonic variables. It is also true that the processing time for these terms is much greater  because the set of possible variable transformations is large. 

\subsection{Easy examples}
\label{easy-sec}

There are some terms that immediately have the form of one of the fundamental integrals and some that can easily be rewritten so that they do. Two examples of the latter are shown in eqs.~(\ref{sh1}, \ref{sh2})
\bea
&&\sumint_{PKQ} \frac{1}{K^2Q^2(K+Q)^2(P+K)^2} \to \sumint_{PKQ}\frac{1}{K^2Q^2(K+Q)^2 P^2} = b_1 I_{\rm bsun} \label{sh1}\\[2mm]
&& \sumint_{PKQ}\frac{1}{P^2Q^2(K+Q)^2(P+K)^2} \to \sumint_{PKQ}\frac{1}{P^2K^2Q^2(P+K+Q)^2} = I_{\rm bball} \label{sh2}\,
\eea
where in (\ref{sh1}) we have used the transformation $P\to-P-K$ and in (\ref{sh2}) we use $K\to -K-P$ and $Q\to -Q$.

\subsection{Simplifying frequency sums}
Some sum-integrals can be simplified using the identities in eq.~(\ref{freq-trick}, \ref{freq-trick-4}) which are well known and can be derived with the change of variable $p=p_0 x$. 
\bea
n\ge 2: ~~ && \sumint \frac{p_0^2}{P^{2n}} = \frac{2n-1-d}{2n-2}\sumint\frac{1}{P^{2(n-1)}} \label{freq-trick} \\[4mm]
n\ge 3: ~~ && \sumint \frac{p_0^4}{P^{2n}} = \frac{(1+d-2n)(3+d-2n)}{4(n-2)(n-1)}\sumint\frac{1}{P^{2(n-2)}}\,. \label{freq-trick-4}
\eea

Most terms that are odd in one of the frequency variables can be rewritten as a sum of terms that are even. One example of this is shown in eq.~(\ref{example2}). 
\bea
&& \sumint_{PKQ} \frac{k_0q_0\delta_{p_0}}{K^2(K+Q)^4 Q^2(P^2+m^2)} \nn \\
&& = -\sumint_{PKQ}\left(\frac{k_0^2 \delta_{q_0}}{P^4 K^2 (P+K)^2 (Q^2+m^2)}+\frac{k_0 p_0 \delta_{q_0}}{P^4 K^2 (P+K)^2 (Q^2+m^2)}\right) \nn \\
&& = -\sumint_{PKQ}\frac{k_0^2 \delta_{q_0}}{P^4 K^2 (P+K)^2 (Q^2+m^2)}+\sumint_{PKQ}\frac{p_0^2 \delta_{q_0}}{2P^4 K^2 (P+K)^2 (Q^2+m^2)} \nn\\[2mm]
&& =  \left(-I_{\rm 2b}+\frac{1}{2} I_{\rm 2a}\right) J_{\rm m1a}\,. \label{example2}
\eea
To get the expression in the second line we perform the shift $Q\to-Q-K$ and then rename variables $Q\leftrightarrow P$. To get the third line we rewrite the last term in the second line $t_{ex}=(t_{ex}+\tilde t_{ex})/2$ where $\tilde t_{ex}$ is constructed from $t_{ex}$ with the transformation $K\to-K-P$. 

The only odd frequency integral that cannot be rewritten in this way is the fundamental integral $I_{\rm 2e}$ (see appendix \ref{int-def-sec}). The reason is that the two frequencies in the numerator are a boson/fermion pair which means that some of the needed transformations are not valid in this case. 

\subsection{Removing tadpoles}
Tadpoles are zero in dimensional regularization which means that for integer $n$
\bea
\sumint_P \frac{\delta_{p_0}}{P^{2n}} = 0 \,.
\eea
In some cases one needs a series of variable transformations to identify tadpoles. An example is shown in eq.~(\ref{applyTAD-example})
\bea
&&\sumint_{PKQ}\frac{ \delta_{p_0}\delta_{k_0}\delta_{q_0}}{P^4((P+K)^2+m^2)((K+Q)^2+m^2)} 
 = \sumint_{PKQ}\frac{ \delta_{p_0}\delta_{k_0}\delta_{q_0}}{P^4(K^2+m^2)(Q^2+m^2)} =0 \,.\nn \\ \label{applyTAD-example}
\eea
To get the second expression we shift variables $Q\to-Q-K$ and then $K\to-K-P$. The final result is zero because the $P$ integral has the tadpole form. 

\subsection{Removing numerator dot products}

There are four fundamental integrals $(H_3, H_4, H_5, H_6)$ that have dot products in their numerators that cannot be removed. In all other cases we must remove any dot products from the numerator. In this section we explain how this is done. 

A simple example is shown in (\ref{oddAterm}). 
\bea
\sumint_{PK} \frac{p_0^{n_1}k_0^{n_2} (P\cdot K)}{P^{2n_3}K^{2n_4}} = \sumint_{PK}\frac{p_0^{n_1+1}k_0^{n_2+1} }{P^{2n_3}K^{2n_4}} \text{~~for~~} \{n_1,n_2,n_3,n_4\} \in \mathbb{Z}\,. \label{oddAterm}
\eea
To get the second expression we use that the integrand is odd under $\vec p \to -\vec p$. The integral is zero if either $n_1+1$ or $n_2+1$ is odd, and if they are both even it can be calculated using  eqs.~(\ref{freq-trick}, \ref{freq-trick-4}). 

In some cases a term with a dot product can be removed by shifting variables and using the symmetry of the resulting integrand. An example is shown in eq.~(\ref{example-OAT1}):
\bea
&& -\sumint_{PKQ} \frac{(K\cdot P) \delta_{p_0}\delta_{k_0}\delta_{q_0} }{K^6 ((P+K)^2+m^2)((K+Q)^2+m^2)} \nn \\
&& = \sumint_{PKQ}  \delta_{p_0}\delta_{k_0}\delta_{q_0} \left(\frac{(P\cdot K)}{K^6 (Q^2+m^2) (P^2+m^2)} + \frac{1}{K^4 (P^2+m^2)(Q^2+m^2)} \right)=0\label{example-OAT1}
\eea
To get the second line we use the transformations $P\to-P-K$ and $Q\to-Q-K$. The first term in the second line is zero because it is odd under $P\to -P$ and the second is zero because the $K$ integral is a tadpole. 

Sometimes a dot product can be removed by writing it in terms of inverse propagators. An example is shown in eq.~(\ref{exampleSUD}):
\bea
&& \sumint_{PKQ} \frac{(P\cdot K)}{P^2K^2(P+K)^2 (P+K+Q)^2} \label{exampleSUD} \\[2mm]
&& = \frac{1}{2}\sumint_{PKQ}\frac{(P+K)^2-P^2-K^2}{P^2K^2(P+K)^2 (P+K+Q)^2} \nn \\[2mm]
&& = \frac{1}{2}\sumint_{PKQ}\left(\frac{1}{P^2K^2 (P+K+Q)^2} - \frac{1}{K^2(P+K)^2(P+K+Q)^2}- \frac{1}{P^2(P+K)^2(P+K+Q)^2}\right) \nn \\
&& = - \frac{1}{2}\sumint_{PKQ}\frac{1}{P^2K^2Q^2} = -\frac{1}{2} b_1^3\,.\nn
\eea
To combine the three terms in the third line we perform the shift $Q\to -Q-P-K$ on all terms and then the additional shifts $P\to-P-K$ on the second term and $K\to -K-P$ on the third. 
This strategy will remove any factor $(P_1\cdot P_2)$ if the denominator has three propagators with the variables $P_1$, $P_2$ and $P_1+P_2$. We will refer to this as a `safe' dot product. If any of these propagators are not present then using this method fails because it will produce an extra term (or terms) with inverse propagators that are not cancelled (which cannot be written in terms of fundamental integrals). 

In some cases the numerator dot product does not have the safe form, but we can use a series of variable transformations to rewrite so that it does, and then use the basic strategy in (\ref{exampleSUD}). An example is shown in  eq.~(\ref{OAT-details}):
\bea
t&=&  \sumint_{PKQ} \frac{P\cdot Q}{Q^2K^2P^2(K+Q)^2(P+K)^2} \nn \\
&=&  -\sumint_{PKQ} \frac{K\cdot P}{2K^2P^2Q^2(P+K)^2(K+Q)^2} \nn \\
&=&  \sumint_{PKQ} \frac{1}{4Q^2(K+Q)^2}\left(\frac{1}{(P+K)^2K^2}+\frac{1}{P^2(P+K)^2}-\frac{1}{P^2K^2}\right) \nn \\
&=&\sumint_{PKQ} \frac{1}{4P^2(P+K)^2(K+Q)^2Q^2} = \frac{1}{4}I_{\rm bball}\,.\label{OAT-details}
\eea
To get the second line we use $t=(t+\tilde t)/2$ where $\tilde t$ is obtained from $t$ using the variable shift $Q\to -Q-K$. To get the third line we rewrite $P\cdot K = ((P+K)^2-P^2-K^2)/2$. To get the last expression we do the variable shift $P\to-P-K$ on the first term in the third line so that it cancels with the third term. 

In soft integrals rotational invariance can be used to remove dot products. There are three basic forms and examples for each are given in eqs.~(\ref{deSING-example}, \ref{deDOUB-example}, \ref{undot1dAterm}). 

The first type has all variables soft and one dot product in the numerator and can be handled as shown in eq.~(\ref{deSING-example}):
\bea
&&  \sumint_{PKQ}  \frac{P\cdot Q \delta_{p_0}\delta_{k_0}\delta_{q_0}}{K^2 (P^2+m^2)(P+K)^2(Q^2+m^2)(K+Q)^2}\label{deSING-example} \\
&& = T^3 \int_{pkq} \frac{p\cdot q }{k^2 (p^2+m^2)(p+k)^2(q^2+m^2)(k+q)^2} \nn \\
&& = T^3 \int_k  \frac{1}{k^2} \left(\int_p \frac{p_i}{(p^2+m^2)(p+k)^2}\right)\left(\int_q \frac{q_i}{(q^2+m^2)(q+k)^2}\right)  \nn \\
&& = T^3 \int_k \frac{1}{k^2} (k_i J_1)(k _iJ_2)   \nn \\
&& = T^3 \int_k \frac{1}{k^2} k_i \left(\int_p\frac{p\cdot k}{k^2(p+k)^2(p^2+m^2)}\right) k_i \int_q\left(\frac{q\cdot k}{k^2(q+k)^2(q^2+m^2)} \right)\nn \\
&& =  T^3 \int_{pkq}\frac{(p\cdot k)(q\cdot k)}{k^4(p+k)^2(p^2+m^2)(q+k)^2(q^2+m^2)} \nn \\
&& = \frac{1}{4}\left(J_{\rm m3b}+2J_{m1a}(J_{\rm m2e}-J_{\rm m2b}) -2J_{\rm m3d} + T^3 m^4\int_{pkq} \frac{1}{(p^2+m^2)(q^2+m^2)k^4 (k+q)^2(p+k)^2}\right)\,.\nn
\eea
In the first line we use the Kronecker deltas to reduce the sum-integrals to three dimensional integrals. In the second line we separate the parts of the integrand that depend on $p$ and $q$. In the third line we use rotational invariance to rewrite these pieces in terms of the soft integrals denoted $J_1$ and $J_2$. In the fourth line we obtain expressions for these integrals by contracting the equations that define them by $k_i$ on both sides, and in the fifth line we rearrange all pieces. The dot products in the numerator of the fifth line can be removed using the usual trick by rewriting them as $\vec p\cdot\vec k = ((p+k)^2-k^2-(p^2+m^2)+m^2)/2$ and $\vec k\cdot\vec q = ((k+q)^2-k^2-(q^2+m^2)+m^2)/2$. This produces a longer expression, but some of the resulting terms are tadpoles and can be immediately removed, and others can be combined using variable transformations. The final result is shown in the sixth line. There are five known soft integrals and one additional term that cancels when all terms are combined. 

The second type has three soft momenta and a squared dot product in the numerator. An example is shown in eq.~(\ref{deDOUB-example}). 
\bea
&&  \sumint_{PKQ} \frac{\delta_{p_0}\delta_{k_0}\delta_{q_0} (P\cdot Q)^2 }{K^4Q^2(K+Q)^2(P^2+m^2)((P+K)^2+m^2)} \label{deDOUB-example} \\
&& = T^3 \int_{pkq} \frac{ (p\cdot q)^2 }{k^4q^2(k+q)^2(p^2+m^2)((p+k)^2+m^2)} \nn\\
&& = T^3 \int_k \frac{1}{k^4}  \int_q  \frac{q_iq_j}{q^2(k+q)^2} \int_p \frac{p_i p_j}{(p^2+m^2)((p+k)^2+m^2)}\nn\,.
\eea
We rewrite the $p$ and $q$ integrals as
\bea
&& \int_q  \frac{q_iq_j}{q^2(k+q)^2} = k_i k_j J_1(K)+\delta_{ij}J_2(k) \\
&& \int_p \frac{p_i p_j}{(p^2+m^2)((p+k)^2+m^2)} = k_i k_j J_3(K)+\delta_{ij}J_4(k)\,.\nn 
\eea
We contract both of these equations with $k_ik_j$ and then make two different equations by contracting with $\delta_{ij}$. This produces four equations we can solve for the scalar integrals $J_i(k)$. Substituting these expressions and rearranging we get
\bea
t_1 &=& \frac{1}{d-2}\int_{pkq} \frac{1}{k^4}\bigg(
\frac{1}{(k+q)^2(p^2+m^2)}
-\frac{m^2}{(k+q)^2(p^2+m^2)((p+k)^2+m^2)}\label{dedoub1} \\[2mm]
&& -\frac{(k\cdot p)^2}{k^2 (k+q)^2(p^2+m^2)((p+k)^2+m^2)}
-\frac{(k\cdot q)^2}{k^2q^2(k+q)^2((p+k)^2+m^2)} \nn \\[2mm]
&& +\frac{m^2(k\cdot q)^2}{k^2 q^2(k+q)^2(p^2+m^2)((p+k)^2+m^2)} 
+\frac{(d-1)(p\cdot k)^2(q\cdot k)^2}{k^4q^2(k+q)^2(p^2+m^2)((p+k)^2+m^2))}
\bigg)\,.\nn
\eea
All of the terms in (\ref{dedoub1})  have at least one safe dot product that can be removed with the basic trick shown in (\ref{exampleSUD}). The resulting expression has at most one numerator dot product in each term. These can be removed using the methods described previously. First we search for terms that can be transformed into safe dot products, using the method in eq.~(\ref{OAT-details}), and remove these safe dot products in the usual way. 
Then we look for tadpole forms and remove them.  
This procedure gives a result that is completely free of numerator dot products but can have many terms. Sometimes a single term expands into over 100 terms when the original squared dot product is removed.
The last step is to take each term and search for a transformation that changes it into one of the fundamental integrals. For this example the final result is
\bea
 \frac{1}{16(d-2)}\left((d-1)J_{\rm m3b}+4J_{\rm m3h}\right)\,.
\eea 

Similar manipulations can be performed on integrals with mixed hard and soft integrals, even when the numerator dot product mixes hard and soft variables. An example is shown in 
eq.~(\ref{undot1dAterm}). 
\bea
 \sumint_{PKQ}\frac{(P\cdot K)^2 q_0^2\delta_{k_0}}{P^6(K^2+m^2)^2 Q^4} &=& T \sumint_{PQ} \frac{q_0^2}{Q^4}\int_k \frac{(p\cdot k)^2 }{P^6(k^2+m^2)^2} \nn \\
&=&  T \sumint_Q\frac{q_0^2}{Q^4} \sumint_P \frac{p_ip_j}{P^6} \underbrace{ \int_k \frac{k_ik_j}{(k^2+m^2)^2} }_{\delta_{ij} J}\nn \\
&=&  T \frac{1}{d-1}\sumint_Q\frac{q_0^2}{Q^4} \sumint_P \frac{(P^2-p_0^2)}{P^6}\int_k \frac{k^2}{(k^2+m^2)^2}. \label{undot1dAterm}
\eea
In the second line we write the $k$ integral in terms of a scalar function denoted $J$. In the third line we use an integral expression for $J$ that was obtained by contracting both sides of the equation that defines it with $\delta_{ij}$. The three integrals in the last line can be easily rewritten in terms of fundamental integrals and give
\bea
 -\frac{1}{16}b_1b_2(d-3)(d-1)J_{\rm m1a}\,.
\eea

\section{Definitions of integrals}
\label{int-def-sec}

There are several different conventions for the integrals that appear in the literature. We mostly use those of \cite{Zhai:1995ac} with the exception of some hard three loop integrals. 
In the original work of Arnold and Zhai \cite{Arnold:1994ps,Arnold:1994eb} the order $g^4$ contribution to the gluon baseball diagram in fig.~\ref{gl-sc-base-fig} was written as $I_{\rm qcd}$ (the symmetry factor, coupling constant and colour factors were extracted). This integral was in turn written in terms of other known hard integrals and another integral called $I_{\rm sqed}$ which came from an earlier calculation of theirs done using scalar QED. Both $I_{\rm qcd}$ and $I_{\rm sqed}$  have three versions corresponding to different possible combinations of bosonic and fermionic lines. The authors of \cite{Zhai:1995ac} used a definition of $I_{\rm qcd}$ that differs by an overall factor. In \cite{Du:2021jai} the authors write $I_{\rm qcd}$ in terms of an integral written $\bar I$ that is equivalent to what is called $I_{\rm sqed}$ by previous authors. All of these integrals can be written in terms of other hard integrals and the three functions $(H_4,H_5,H_6)$.  In our calculation we use exclusively the $H$ functions which are produced directly by our program. 
For all other integrals we use the notation of \cite{Zhai:1995ac}. 
Integrals with all propagators soft are denoted $J$. 
For each integral with $m$ in the subscript (and a gluon mass in the integrand) there is a corresponding definition with $m\to M$. 
Some of the integrals defined below are not independent and a list of identities that relate them is given in section \ref{identities-sec}. Results for some of the independent integrals are given in section \ref{int-results-sec}, for others we give a reference to the equation in a previous paper where the result can be found. In section \ref{results-diags-sec} we give the results for each diagram in terms of the integrals listed in this section. 
\bea
&& \hspace*{-1cm} \text{hard one loop integrals} \nn \\
&& b_0 = \sumint_P \log(P^2) \nn \\
&& b_1=\sumint_P \frac{1}{P^2} \label{1-hard} \\
&& b_2=\sumint_P \frac{1}{P^4} \nn \\
&& f_0 = \sumint_{\{P\}} \log(P^2) \nn \\
&& f_1=\sumint_{\{P\}} \frac{1}{P^2} \nn \\
&& f_2=\sumint_{\{P\}} \frac{1}{P^4} \nn \\[4mm]
%%%%%%%%%%%%%%%%%%%%%%%
&& \hspace*{-1cm} \text{hard two loop integrals} \label{2-hard} \\
&& I_{\rm 2a} = \sumint_{PK}\frac{p_0^2}{P^4K^2(P+K)^2}\nn \\
&& I_{\rm 2b} = \sumint_{PK}\frac{p_0^2}{P^2K^4(P+K)^2}\nn \\
&& I_{\rm 2c} = \sumint_{P\{K\}}\frac{p_0^2}{P^4K^2(P+K)^2}\nn \\
&& I_{\rm 2d} = \sumint_{\{P\}K}\frac{p_0^2}{P^2K^4(P+K)^2}\nn \\
&& I_{\rm 2e} =  \sumint_{\{P\}K}\frac{p_0 k_0}{P^4K^2(P+K)^2}\nn \\
&& I_{\rm bsun} = \sumint_{PK}\frac{1}{P^2K^2(P+K)^2} \nn\\
&& I_{\rm fsun} = \sumint_{P\{K\}}\frac{1}{P^2K^2(P+K)^2} \nn\\
&& A_{\rm 1b}= \sumint_{PK}\frac{\delta_{p_0}}{P^2K^2(P+K)^2}\nn \\
&& A_{\rm 2b}= - \sumint_{PK} \frac{\delta_{p_0}k_0^2}{P^4 K^2 (P+K)^2}\nn \\
&& A_{\rm 1f}= \sumint_{P\{K\}}\frac{\delta_{p_0}}{P^2K^2(P+K)^2}\nn \\
&& A_{\rm 2f}= - \sumint_{P\{K\}} \frac{\delta_{p_0}k_0^2}{P^4 K^2 (P+K)^2}\nn \\[4mm]
%%%%%%%%%%%%%%%%%%%%%%%%%%%%%%%%%%%%%%%%%%
&& \hspace*{-1cm} \text{hard/soft two loop integrals} \label{2-soft-hard} \\
&& F_{\rm m2b} = \sumint_{PK}\frac{\delta_{p_0}}{(P^2+m^2)K^2(P+K)^2} \nn \\
&& \tilde F_{\rm m2b} = \sumint_{PK}\frac{\delta_{p_0} k_0^2 \Delta_m(P)}{K^2(P+K)^2} \nn\\
&& F_{\rm m2f} = \sumint_{P\{K\}}\frac{\delta_{p_0}}{(P^2+m^2)K^2(P+K)^2} \nn \\
&& \tilde F_{\rm m2f} = \sumint_{P\{K\}}\frac{\delta_{p_0} k_0^2 \Delta_m(P)}{K^2(P+K)^2} \nn\\[4mm]
%%%%%%%%%%%%%%%%%%%%%%%%%%%%%%%%%%%%%%%%%%%%%
&& \hspace*{-1cm} \text{hard three loop integrals} \label{3-hard} \\
&& I_{\rm bball} = \sumint_{PKQ}\frac{1}{P^2K^2Q^2(P+K+Q)^2} \nn\\
&& I_{\rm fball} = \sumint_{P\{K\} Q}\frac{1}{P^2K^2Q^2(P+K+Q)^2} \nn\\
&& I_{\rm ffball} = \sumint_{\{P\}\{K\}\{Q\}}\frac{1}{P^2K^2Q^2(P+K+Q)^2} \nn\\
%%%%%%%%%%%%%%%%%
&& H_3 = \sumint_{P\{K\}Q} \frac{(P\cdot Q)}{P^2K^2Q^2(P+K)^2(K+Q)^2} \nn \\
&& H_4 =\sumint_{P\{KQ\}} \frac{(Q\cdot K)^2}{P^4Q^2K^2(P+Q)^2(P+K)^2}  \nn\\
&& H_5 = \sumint_{PQK} \frac{(Q\cdot K)^2}{P^4Q^2K^2(P+Q)^2(P+K)^2} \nn\\
&& H_6 = \sumint _{P\{K\}Q} \frac{(Q\cdot K)^2}{P^4Q^2K^2(P+Q)^2(P+K)^2} \nn\\[4mm]
%%%%%%%%%%%%%%%%%%%%%%%%%%%%%%%%
&&\hspace*{-1cm} \text{soft one loop integrals} \label{1-soft} \\
&& J_{\rm m0} = \sumint_P \delta_{p_0} \log(P^2+m^2) \nn \\
&& J_{\rm m1a} = \sumint_P\frac{\delta_{p_0}}{P^2+m^2} \nn \\[4mm]
&& \hspace*{-1cm} \text{soft two loop integrals} \label{2-soft} \\
&& J_{\rm m2a} = \sumint_{PK} \frac{\delta_{p_0} \delta_{k_0}}{(P^2+m^2)(K^2+m^2)(P+K)^2} \nn \\
&& J_{\rm m2b} = \sumint_{PK} \frac{\delta_{p_0} \delta_{k_0}}{(P^2+m^2)K^2(P+K)^2} \nn \\
&& J_{\rm m2c} = m^2\sumint_{PK} \frac{\delta_{p_0} \delta_{k_0}}{(P^2+m^2)(K^2+m^2)(P+K)^4} \nn \\
&& J_{\rm m2d} = m^2\sumint_{PK} \frac{\delta_{p_0} \delta_{k_0}}{(P^2+m^2)^2(K^2+m^2)(P+K)^2} \nn \\
&& J_{\rm m2e} = m^2\sumint_{PK} \frac{\delta_{p_0} \delta_{k_0}}{(P^2+m^2)K^2(P+K)^4} \nn \\
&& J_{\rm m2f} = m^2\sumint_{PK} \frac{\delta_{p_0} \delta_{k_0}}{(P^2+m^2)^2K^2(P+K)^2} \nn \\[4mm]
%%%%%%%%%%%%%%%%%%%%%%%%%%%%%%%%%%%%%%%%%%%%
&& \hspace*{-1cm} \text{soft three loop integrals} \label{3-soft} \\
&& J_{\rm m3a} = \sumint_{PKQ} \frac{\delta_{p_0} \delta_{k_0}\delta_{q_0}}{(P^2+m^2)(K^2+m^2)(Q^2+m^2)((P+K+Q)^2+m^2)} \nn \\
&& J_{\rm m3b} = \sumint_{PKQ} \frac{\delta_{p_0} \delta_{k_0}\delta_{q_0}}{(P^2+m^2)K^2(Q^2+m^2)(P+K+Q)^2} \nn \\
&& J_{\rm m3c} = m^2\sumint_{PKQ} \frac{\delta_{p_0} \delta_{k_0}\delta_{q_0}}{(P^2+m^2)K^2(Q^2+m^2)((P+K)^2+m^2)((K+Q)^2+m^2)} \nn \\
&& J_{\rm m3d} = m^2\sumint_{PKQ} \frac{\delta_{p_0} \delta_{k_0}\delta_{q_0}}{P^2K^2Q^2((P+K)^2+m^2)((K+Q)^2+m^2)} \nn \\
&& J_{\rm m3e} = m^2\sumint_{PKQ} \frac{\delta_{p_0} \delta_{k_0}\delta_{q_0}}{(P^2+m^2)(K^2+m^2)(Q^2+m^2)(P+K)^2(K+Q)^2} \nn \\
&& J_{\rm m3f} = m^4\sumint_{PKQ} \frac{\delta_{p_0} \delta_{k_0}\delta_{q_0}}{(P^2+m^2)(K^2+m^2)^2(Q^2+m^2)(P+K)^2(K+Q)^2} \nn \\
&& J_{\rm m3g} = m^4\sumint_{PKQ} \frac{\delta_{p_0} \delta_{k_0}\delta_{q_0}}{(P^2+m^2)K^4(Q^2+m^2)((P+K)^2+m^2)((K+Q)^2+m^2)} \nn \\
&& J_{\rm m3h} = m^2\sumint_{PKQ} \frac{\delta_{p_0} \delta_{k_0}\delta_{q_0}}{P^2K^2(Q^2+m^2)(P+K)^2((K+Q)^2+m^2)} \nn \\
&& J_{\rm m3i} = m^4\sumint_{PKQ} \frac{\delta_{p_0} \delta_{k_0}\delta_{q_0}}{(P^2+m^2)K^2(Q^2+m^2)((P+K)^2+m^2)((K+Q)^2+m^2)(P+K+Q)^2} \nn \\
&& J_{\rm m3j} = \sumint_{PKQ} \frac{\delta_{p_0} \delta_{k_0}\delta_{q_0}(P\cdot Q)}{(P^2+m^2)(K^2+m^2)(Q^2+m^2)(P+K)^2(K+Q)^2} \nn \\[4mm]
&& \hspace*{-1cm} \text{soft three loop integrals that mix $m$ and $M$} \label{3-soft-mixed} \\
&& J_{\rm 3m} = \sumint_{PKQ} \frac{\delta_{p_0} \delta_{k_0}\delta_{q_0}}{(P^2+m^2)K^2(Q^2+M^2)((P+K)^2+m^2)((K+Q)^2+M^2)} \nn \\
&& J_{\rm 3n} = \sumint_{PKQ} \frac{\delta_{p_0} \delta_{k_0}\delta_{q_0}}{(P^2+M^2)(Q^2+m^2)((P+K)^2+m^2)((K+Q)^2+M^2)} \nn \\
&& J_{\rm 3p} = \sumint_{PKQ} \frac{\delta_{p_0} \delta_{k_0}\delta_{q_0}}{(P^2+m^2)K^4(Q^2+M^2)((P+K)^2+m^2)((K+Q)^2+M^2)}\,. \nn 
\eea

%%%%%%%%%%%%%%%%%%%%%%%%%%%%%%%%%%%%%%%%%%%%%%%%%%%%%%%%%%%%%%%%%%%%%%
\section{Identities connecting fundamental integrals}
In this section we give a list of the identities we use that relate some of the fundamental integrals in section \ref{int-def-sec}. The method to derive them is explained in \cite{Zhai:1995ac}. For each identity that relates functions with subscript $m$ there is  a corresponding identity with subscript $M$. 
\label{identities-sec}
\bea
\label{identities}
&& F_{\text{m2b}} = A_{\text{1b}}+b_2 J_{\text{m1a}}+J_{\text{m2b}} \nn \\
&& F_{\text{m2f}} =  A_{\text{1f}}+f_2 J_{\text{m1a}} \nn \\
&& \tilde{F}_{\text{m2b}} = A_{\text{2b}} m^2-\frac{1}{12} (d-5) b_2
   J_{\text{m1a}} m^2-\frac{1}{2} (d-3) b_1 J_{\text{m1a}} \nn \\
&& \tilde{F}_{\text{m2f}} = A_{\text{2f}} m^2-\frac{1}{12} (d-5) f_2
   J_{\text{m1a}} m^2-\frac{1}{2} (d-3) f_1 J_{\text{m1a}} \nn \\
&& J_{\text{m2a}} = -\frac{(d-3) J_{\text{m1a}}^2}{2 (d-4) m^2}\nn \\
&& J_{\text{m2b}} = \frac{11 J_{\text{m3b}}-3 d J_{\text{m3b}}}{6 J_{\text{m1a}}-2 d J_{\text{m1a}}}-\frac{J_{\text{m3d}}}{2 J_{\text{m1a}}} \nn \\
&& J_{\text{m2c}} = \frac{J_{\text{m2d}}}{d-6} \nn \\
&& J_{\text{m2d}} = -\frac{1}{2} (d-4) J_{\text{m2a}} \nn \\
&& J_{\text{m2e}} =  (d-4) J_{\text{m2b}}\nn \\
&& J_{\text{m2f}} =  \frac{3 J_{\text{m2b}}}{d+1}-\frac{J_{\text{m2e}}}{d+1}\nn \\
&& J_{\text{m3c}} = \frac{(d-3) J_{\text{m1a}} J_{\text{m2a}}}{d-5}+\frac{(11-3 d) J_{\text{m3a}}}{4(d-5)} \nn \\
&& J_{\text{m3e}} = \frac{(d-3)^2 J_{\text{m1a}}^3}{4 (d-4)^2 m^2}+\frac{(3 d-11)J_{\text{m3b}}}{4 (d-4)} \nn \\
&& J_{\text{m3f}} = \frac{(3 d-13) J_{\text{m3e}}}{4d-18}-\frac{(d-3) J_{\text{m1a}} J_{\text{m2d}}}{2 d-9} \nn \\
&& J_{\text{m3g}} = \frac{(d-3) J_{\text{m1a}} J_{\text{m2c}}}{d-7}+\frac{(13-3 d) J_{\text{m3c}}}{4(d-7)} \nn \\
&& J_{\text{m3h}} = \frac{(11-3 d) J_{\text{m3b}}}{8 d-36}\nn \\
&& J_{\text{m3i}} = \frac{(13-3 d) J_{\text{m3c}}}{4 (d-5)}+\frac{(3 d-13) J_{\text{m3e}}}{2(d-5)}+\frac{J_{\text{m3f}}}{d-5} \nn \\
&& J_{\text{m3j}} = -\frac{J_{\text{m1a}}^3}{4m^2}-J_{\text{m2a}} J_{\text{m1a}}+\frac{J_{\text{m2b}}J_{\text{m1a}}}{2}+\frac{J_{\text{m3b}}}{4}+\frac{J_{\text{m3d}}}{4}-J_{\text{m3e}} \,.\nn 
\eea

%%%%%%%%%%%%%%%%%%%%%%%%%%%%%%%%%%%%%%%%%%%%%%%%%%%%%%%%%%
\section{Results for fundamental integrals}
\label{int-results-sec}

In this appendix we list the places that results for some of  the known integrals in appendix \ref{int-def-sec} can be found. We note that the papers we reference below are not necessarily the ones in which the integrals were originally calculated, but a place where the result can be easily extracted. 

In ref.~\cite{Zhai:1995ac} one can find the results for
the hard one loop integrals (\ref{1-hard}) that do not have logarithms in the integrand in eq.~(B2);
the hard two loop integrals (\ref{2-hard}) are (B3, B6);
and the soft one loop integral (\ref{1-soft}) and the soft three loop integrals $J_{m3a} \dots J_{m3j}$ (\ref{3-soft}) are given in eq.~(B7). 
The logarithmic integrals $b_0$ and $f_0$ are found in eqs.~(4.6, 4.10) of \cite{Du:2021jai}, and $J_{m0}$ is given in eq.~(\ref{J0tilde}). 
The hard three loop integrals (\ref{3-hard}) are given in eqs.~(A14-A20) of ref.~\cite{Andersen:2021bgw}. 
The remaining hard integrals $(A_{1b},A_{2b},A_{1f},A_{2f})$ can be extracted from \cite{Arnold:1994ps,Arnold:1994eb} and are
\bea
&& A_{1b} = \frac{T^2}{16 \pi ^2} \left(-\frac{1}{4 \epsilon
   }-\frac{1}{2}+ \log \left(\frac{2 T}{\bar\mu}\right)\right) \nn \\
&& A_{2b} = -\frac{T^2}{128 \pi ^2} \nn \\
&& A_{1f} = -\frac{T^2 \log (2)}{16 \pi ^2} \nn \\
&& A_{2f} = 0\,.
\eea
The only new integrals we need are the soft three loop integrals (\ref{3-soft-mixed}) which are
\bea
&&J_{\rm 3m} = \frac{T^3}{128 \pi ^3} 
\left(
\frac{\log \left(\frac{m+M}{m}\right)}{M}  + [m\leftrightarrow M]
\right) 
\nn \\
&&J_{\rm 3n} = -\frac{T^3}{16\pi^3}\left(
\frac{m}{8 \epsilon }+m
-\frac{1}{2} m \log \left(\frac{m+M}{m}\right)+\frac{3}{4} m \log
   \left(\frac{\bar\mu}{m}\right)-\frac{3}{4} m \log (2) + [m\leftrightarrow M]
\right)\nn \\[3mm]
&&J_{\rm 3p} = -\frac{T^3}{3072 m^3 M^3 \pi^3}\left(
m^2 M+m^3 \log \left(\frac{(m+M)^2}{m^2}\right)+ [m\leftrightarrow M]
\right) \,.
\eea
We also give results for the gluon and scalar thermal masses which are easy to calculate. 
\bea
m^2 &=& g^2 C_A (d-2)\left( (D+4) b_1 -8f_1\right)\nn \\
M^2 &=& g^2 C_A \left((D+4)b_1 -8f_1\right)
\,.
\eea

%%%%%%%%%%%%%%%%%%%%%%%%%%%%%%%%%%%%%%%%%%%%%%%%%%%%%%%%%%
\section{Results for individual diagrams}
\label{results-diags-sec}

\begin{figure}[H]
\begin{centering}
%trim [left bottom right top]
\includegraphics[trim={0cm 0cm 0cm 0cm},clip,scale=0.68]{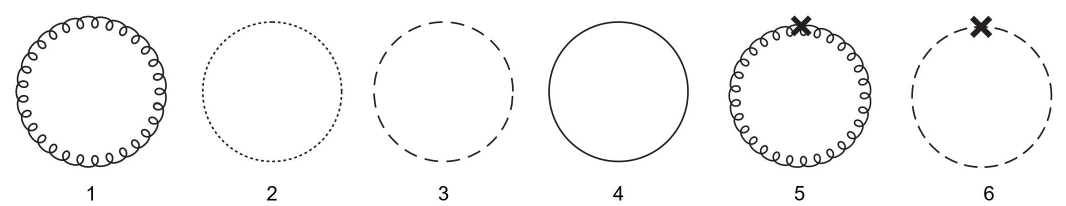}
\caption{One loop graphs and one loop counterterms.\label{1loop-fig}}
\end{centering}
\end{figure}

\begin{figure}[H]
\begin{centering}
\hspace*{-8cm} \includegraphics[scale=1.6]{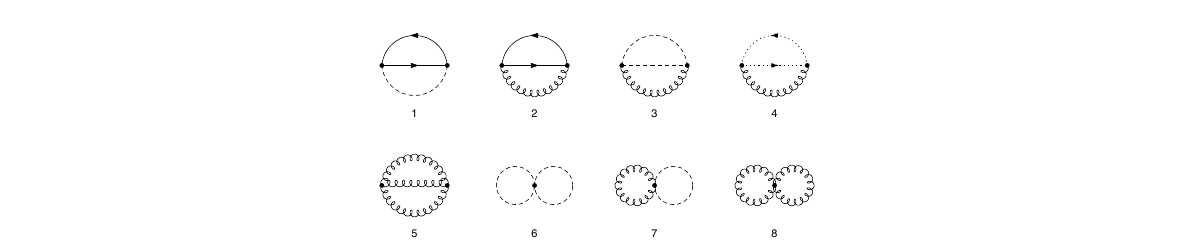}
\caption{Two loop graphs.\label{2loop-fig}}
\end{centering}
\end{figure}

\begin{figure}[H]
\begin{center}
\includegraphics[trim={3cm 3cm 0cm 3cm},clip,scale=1.0]{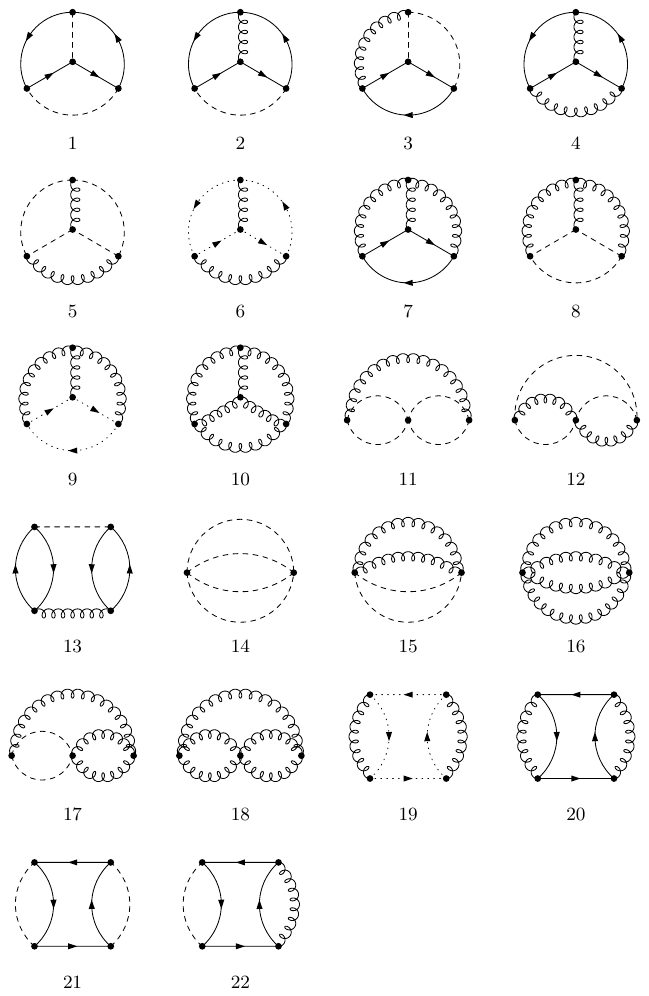}
\caption{Three loop graphs excluding gluon and scalar baseballs. \label{3loop-fig}}
\end{center}
\end{figure}

\begin{figure}[H]
\begin{center}
%trim [left bottom right top]
\includegraphics[scale=0.54]{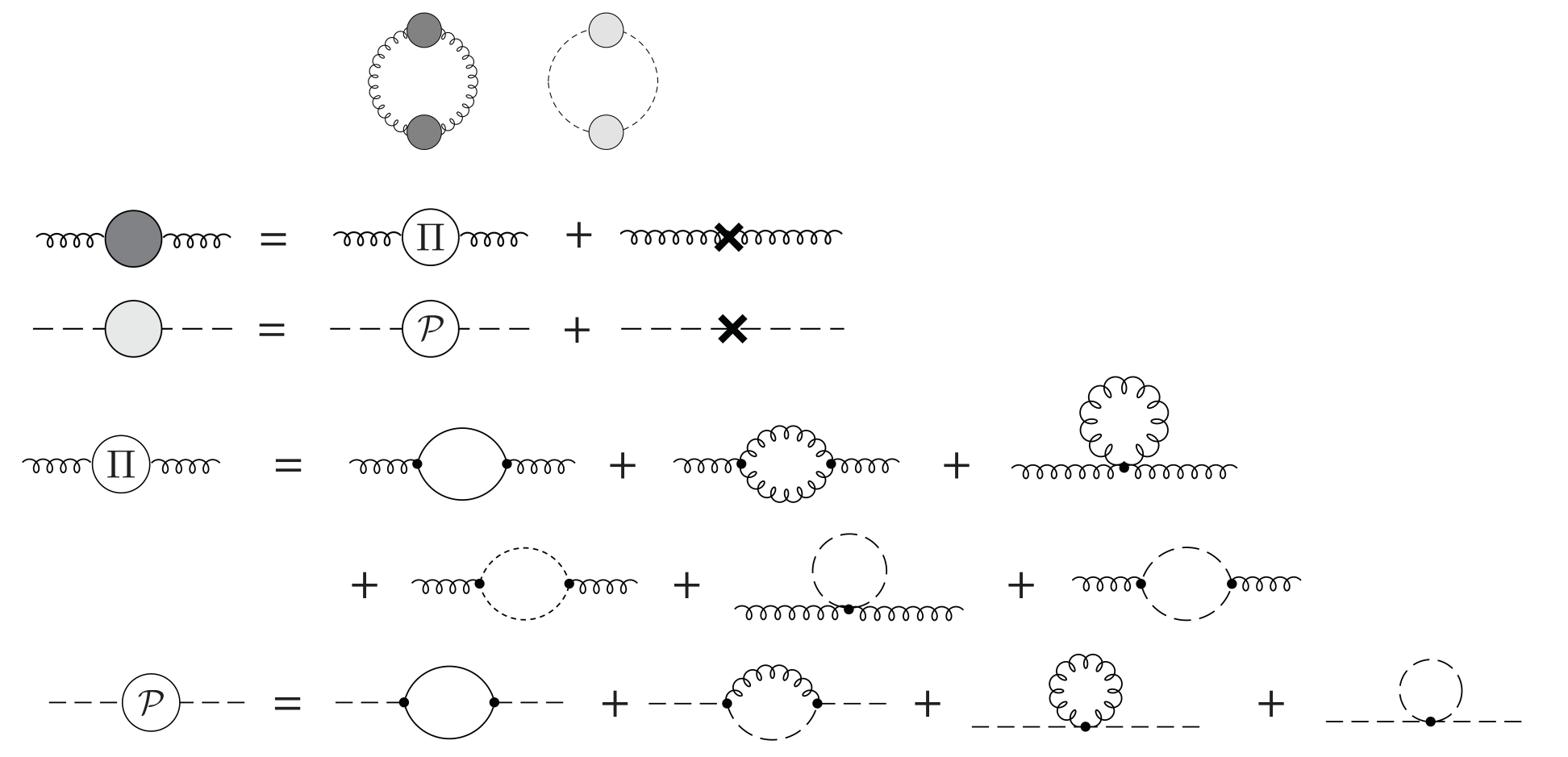}
\caption{Three loop gluon and scalar baseballs (top), and the gluon self energy insertion (dark blob) and scalar self energy insertion (light blob).\label{gl-sc-base-fig}}
\end{center}
\end{figure}

In this section we give results for the individual diagrams. The free energy is divided into contributions from one loop diagrams including those with counterterm insertions (fig.~\ref{1loop-fig}), two loop diagrams (fig.~\ref{2loop-fig}), three loop diagrams that don't involve gluon or scalar baseballs (fig.~\ref{3loop-fig}), and three loop gluon and scalar baseballs with their counterterms contributions (fig.~\ref{gl-sc-base-fig}). The integrals $(I_{2a}, A_{2f}, I_{\rm bsun}, I_{\rm fsun})$ are zero and we have removed them in the expressions below. 
For the one loop diagrams in fig.~\ref{1loop-fig} we define $F^1=-{\cal F}_1/d_A$ and for the two loop contributions to the free energy we write $F^2=-{\cal F}_2/(g^2C_A d_A)$. 
For all three loop contributions we extract a factor $-g^4C_A d_A$. The diagrams without gluon or scalar baseballs are written as $F^3=-{\cal F}_3/(g^4C_A^2 d_A)$ and for gluon and scalar baseballs we write $B_{\rm gl} =-{\cal F}_3/(g^2 C_A^2 d_A)$ and $B_{\rm sc} =-{\cal F}_3/(g^2 C_A^2 d_A)$. 
We use subscripts for each diagram in the order they are drawn in figures \ref{1loop-fig} and \ref{2loop-fig} and \ref{3loop-fig}. For the gluon and scalar baseballs in fig. \ref{gl-sc-base-fig} each amplitude is also divided into contributions from two boson self-energy insertions, one boson and one fermion self-energy insertion, and two fermion self-energy insertions, which are denoted with superscripts.

The results for each diagram are:
\bea
&&\text{one loop diagrams} \nn \\
&& F^1_1 =-\frac{1}{2}(D b_0+J_{m0}) \nn \\
&& F^1_2 = b_0 \nn \\
&& F^1_3 = -3(b_0+ J_{M0})\nn \\
&& F^1_4 = 4f_0 \nn \\
&& F^1_5 = \frac{1}{2}m^2 J_{m1a} \nn \\
&& F^1_6 = 3M^2 J_{M1a} \nn \\[4mm]
%%%%%%%%%%%%%%%%%%%%%%%%%%%%%
&&\text{two loop diagrams} \nn \\
&& F^2_1 = 12 \left(2 b_1 f_1-f_1^2+M^2 F_{\text{M2f}}\right) + 24 f_1 J_{\text{M1a}} \nn \\
&& F^2_2 = -8 \tilde{F}_{\text{m2f}}+2 (D-2) f_1 \left(2 b_1-f_1\right)+2 m^2 F_{\text{m2f}}
+ 4 f_1 J_{\text{m1a}} \nn \\
&& F^2_3  = 6 \tilde{F}_{\text{m2b}}+\frac{9 b_1^2}{2}-6 M^2 F_{\text{M2b}}
+ 3 b_1 J_{\text{M1a}}+6 M^2 \left(J_{\text{M2b}}-J_{\text{M2a}}\right)
-\frac{3 J_{\text{M1a}}^2}{2} \nn \\
&& F^2_4 = \frac{1}{4} \left(-2 \tilde{F}_{\text{m2b}}-b_1^2\right) \nn \\
&& F^2_5 = \left(D-\frac{3}{2}\right) \tilde{F}_{\text{m2b}}+\frac{3}{4} b_1^2 (D-1)
- m^2 F_{\text{m2b}}
+ \frac{b_1 J_{\text{m1a}}}{2}+m^2
   \left(J_{\text{m2b}}-J_{\text{m2a}}\right)
-\frac{J_{\text{m1a}}^2}{4} \nn \\
&& F^2_6 = -\frac{15 b_1^2}{2} -15 b_1 J_{\text{M1a}} -\frac{15 J_{\text{M1a}}^2}{2} \nn \\
&& F^2_7 = -3 b_1^2 D -3 b_1 \left(D J_{\text{M1a}}+J_{\text{m1a}}\right) -3 J_{\text{m1a}}J_{\text{M1a}} \nn \\
&& F^2_8 = -\frac{1}{4} b_1^2 (D-1) D -\frac{1}{2} b_1 (D-1) J_{\text{m1a}}\nn \\[4mm]
&&\text{three loop diagrams except bosonic baseballs} \nn \\
%%%%%%%%%%%%%%%%%%%%%%%%%%%%%%
&& F^3_1 = 24 \text{I}_{\text{fball}}-12 \text{I}_{\text{ffball}} \nn \\
&& F^3_2 = 12 (D-2) \text{I}_{\text{fball}}-6 (D-3) \text{I}_{\text{ffball}}
+ 48 \text{I}_{\text{2e}} J_{\text{m1a}} \nn \\
&& F^3_3 = 12 \text{I}_{\text{fball}} + 48 J_{\text{m1a}}
   \left(\text{I}_{\text{2c}}+\text{I}_{\text{2e}}\right) \nn \\
&& F^3_4 = \frac{1}{2} (D-2) \left(2 (D-4) \text{I}_{\text{fball}}-(D-6)
   \text{I}_{\text{ffball}}\right) + 8 (D-2) \text{I}_{\text{2e}} J_{\text{m1a}} \nn \\
&& F^3_5 = \frac{15 \text{I}_{\text{bball}}}{8} + \frac{3}{8} \left(3 J_{\text{M3a}}+2
   \left(J_{\text{M3b}}+8 J_{\text{M3c}}+J_{\text{M3d}}-16 J_{\text{M3e}}+8
   J_{\text{M3i}}\right)\right) \nn \\
&& ~~~~~~ + \frac{3}{2} J_{\text{M1a}} \left(J_{\text{M2b}}-4
   J_{\text{M2a}}\right) -\frac{3 J_{\text{M1a}}^3}{4 M^2} \nn \\
&& F^3_6 = -\frac{\text{I}_{\text{bball}}}{32} \nn \\
&& F^3_7 = 2 (D-2) \text{I}_{\text{fball}} + 8 J_{\text{m1a}} \left((D-3) \text{I}_{\text{2c}}+2
   \text{I}_{\text{2d}}+(D-2) \text{I}_{\text{2e}}+b_2 (d-3) f_1\right) \nn \\
&& F^3_8 = \frac{15 \text{I}_{\text{bball}}}{8} + \frac{3}{8} \left(-16 J_{\text{m1a}} \left(2
   \text{I}_{\text{2b}}+b_1 b_2 (d-3)\right)+5 J_{\text{M3b}}+J_{\text{M3d}}-16
   J_{\text{M3e}}\right) \nn \\
&& ~~~~~~ + \frac{3}{4} J_{\text{M1a}} \left(J_{\text{M2b}}-4
   J_{\text{M2a}}\right)-\frac{3 J_{\text{M1a}}^3}{8 M^2} \nn \\
&& F^3_9 = -\frac{\text{I}_{\text{bball}}}{16} + \frac{1}{4} J_{\text{m1a}} \left(2
   \text{I}_{\text{2b}}+b_1 b_2 (d-3)\right) \nn \\
&& F^3_{10} = \frac{1}{32} (20 D-23) \text{I}_{\text{bball}}-\frac{1}{4} (4 D-7) J_{\text{m1a}}
   \left(2 \text{I}_{\text{2b}}+b_1 b_2 (d-3)\right) \nn \\
&& ~~~~~~  +\frac{3 J_{\text{m3a}}}{16}+\frac{7
   J_{\text{m3b}}}{16}+J_{\text{m3c}}+\frac{3 J_{\text{m3d}}}{16}-3
   J_{\text{m3e}}+J_{\text{m3i}}+\frac{3}{8} J_{\text{m1a}} \left(J_{\text{m2b}}-4
   J_{\text{m2a}}\right) -\frac{3 J_{\text{m1a}}^3}{16 m^2} \nn \\
&& F^3_{11} = 0 \nn \\
&& F^3_{12} = -\frac{81 \text{I}_{\text{bball}}}{8} -\frac{9}{8} \left(9
   J_{\text{M3b}}+J_{\text{M3d}}-16 J_{\text{M3e}}\right) + J_{\text{M1a}} \left(9
   J_{\text{M2a}}-\frac{9 J_{\text{M2b}}}{4}\right) + \frac{9 J_{\text{M1a}}^3}{8 M^2} \nn \\
&& F^3_{13}=0 \nn \\
&& F^3_{14} = \frac{45 \text{I}_{\text{bball}}}{8} + \frac{45 J_{\text{M3a}}}{8} \nn \\
&& F^3_{15} = \frac{9}{4} \left(D \text{I}_{\text{bball}}+J_{3n}\right)
+ \frac{9}{4} (D-1)
   J_{\text{M3b}} \nn \\
&& F^3_{16} = \frac{3}{16} (D-1) D \text{I}_{\text{bball}} + \frac{3}{8} (D-1) J_{\text{m3b}} \nn \\
&& F^3_{17} = 0 \nn \\
&& F^3_{18} = -\frac{27}{16} (D-1) \text{I}_{\text{bball}} + 3 J_{\text{m3e}}-\frac{3}{16} \left(9
   J_{\text{m3b}}+J_{\text{m3d}}\right) + \frac{3}{8} J_{\text{m1a}} \left(4
   J_{\text{m2a}}-J_{\text{m2b}}\right) + \frac{3 J_{\text{m1a}}^3}{16 m^2} \nn \\
&& F^3_{19} = -\frac{\text{I}_{\text{bball}}}{8} \nn \\
&& F^3_{20} = -2 (D-2)^2 \left(f_2 \left(b_1-f_1\right){}^2+\text{I}_{\text{fball}}-2
   H_3 \right)
+ 4 (D-2) J_{\text{m1a}} \left(-4 \text{I}_{\text{2e}}-(d-4) f_2
   \left(b_1-f_1\right)\right) \nn \\
&& F^3_{21} = -72 \left(f_2 \left(b_1-f_1\right){}^2+\text{I}_{\text{fball}}-2 H_3\right)+ 144
   f_2 \left(f_1-b_1\right) J_{\text{M1a}} \nn \\
&& F^3_{22} = -24 (D-2) \left(f_2 \left(b_1-f_1\right){}^2+\text{I}_{\text{fball}}-2
   H_3\right) \nn \\
&& ~~~~~~~ + 24 \left(-4 \text{I}_{\text{2e}} J_{\text{m1a}}-b_1 f_2 \left((d-4)
   J_{\text{m1a}}+(D-2) J_{\text{M1a}}\right) +f_1 f_2 \left((d-4) J_{\text{m1a}}+(D-2)
   J_{\text{M1a}}\right)\right) \nn \\[4mm]
&&\text{three loop gluon and scalar baseballs} \nn \\
%%%%%%%%%%%%%%%%%%%%%%%%%%%%%%%%%%%%%%%%%%%%%%%
&& B_{\rm gl}^{\rm bb} =  b_1 (d-2) (D+4) A_{\text{1b}}-b_1 (d-2) (d+4) (D+4) A_{\text{2b}}+\frac{3 m^2
   J_{\text{3m}}}{d-2}+\frac{3 M^2 J_{\text{3m}}}{d-2}+\frac{3
   J_{\text{3n}}}{4 (d-2)} \nn \\
&& ~~~~~~ +\frac{12 m^2 M^2 J_{\text{3p}}}{d-2} +\frac{1}{4}
   b_1^2 b_2 (D-4) (D+4)^2-\frac{1}{16} \left(D^2+64\right) \text{I}_{\text{bball}}+(D+4)^2
   H_5 \nn \\
&& ~~~~~~ +2 (D+4) \text{I}_{\text{2b}} J_{\text{m1a}}-\frac{b_1^2 (d-3) (d-2)^2 (D+4) (D-d)
   J_{\text{m1a}}}{4 m^2} \nn \\ 
&& ~~~~~~ -\frac{1}{24} b_2 b_1 (D+4) ((d ((d-8) d+5)+38) D -d (d ((d-8)
   d+17)-34)-72) J_{\text{m1a}} \nn \\
&& ~~~~~~ +3 b_2 b_1 (D-2) (D+4) J_{\text{M1a}}+\frac{(-(d-4) D+d (8
   d-53)+88) J_{\text{m3b}}}{8 (d-4) (2 d-9)} \nn \\
&& ~~~~~~ +\frac{3 (-4 d+D+5) J_{\text{M3b}}}{4
   (d-2)}+\frac{3 (-4 d+D+5) J_{\text{M3h}}}{d-2}+\frac{(d-3) J_{\text{m3a}}}{4 (d-7)
   (d-5)}+\frac{9 J_{\text{M3a}}}{4 (d-2)}+\frac{18 J_{\text{M3c}}}{d-2} \nn \\
&& ~~~~~~ +\frac{36
   J_{\text{M3g}}}{d-2} + \frac{(d-3) (d (2 d-33)+113) J_{\text{m1a}} J_{\text{m2a}}}{4 (d-7) (d-5) (2 d-9)}-\frac{3
   (d-3) J_{\text{m1a}} J_{\text{M2a}}}{2 (d-6)} \nn \\
&& ~~~~~~ -\frac{3 (d-3) (4 d-23) J_{\text{M1a}}
   J_{\text{m2a}}}{2 (d-6)} -\frac{9 (d-3) J_{\text{M1a}} J_{\text{M2a}}}{d-6} \nn \\
&& ~~~~~~ -\frac{3 (d-3) J_{\text{m1a}}^2 J_{\text{M1a}}}{2 m^2}-\frac{9 (d-3) J_{\text{m1a}}
   J_{\text{M1a}}^2}{2 m^2}-\frac{(d-3) (d (d (2 d-21)+72)-84) J_{\text{m1a}}^3}{8 (d-4)^2
   (2 d-9) m^2} \nn\\
%%%%%%%%%%%%%%%%%%%%%%%%%%%%%%%%%%%%%%%%%%%%%%%%
&& B_{\rm gl}^{\rm bf} = -8 (d-2) f_1 A_{\text{1b}}-2 b_1 (d-2) (D+4) A_{\text{1f}}+8 (d-2) (d+4) f_1 A_{\text{2b}}-4
   b_1 b_2 \left(D^2-16\right) f_1 \nn \\
&& ~~~~~~ -2 (D-8) \text{I}_{\text{fball}}-16 (D+4) \text{H}_6 -8 (D-3) \text{I}_{\text{2c}} J_{\text{m1a}}-16 \text{I}_{\text{2d}} J_{\text{m1a}}\nn \\
&& ~~~~~~ +\frac{2
   b_1 (d-3) (d-2)^2 f_1 (D-d) J_{\text{m1a}}}{m^2}+\frac{1}{3} b_2 f_1 ((d ((d-8) d+5)+38)
   D \nn \\
&& ~~~~~~ -d (d ((d-8) d+17)-34)-72) J_{\text{m1a}}-24 b_2 (D-2) f_1 J_{\text{M1a}} \nn \\
&& B_{\rm gl}^{\rm ff}  = 16 (d-2) f_1 A_{\text{1f}}+16 b_2 (D-4) f_1^2+4 (D-4) \text{I}_{\text{ffball}}+64 \text{H}_4\nn\\
%%%%%%%%%%%%%%%%%%%%%%%%%%%%%%%%%%%%%%%%%%%%%%%%%%%
&& B_{\rm sc}^{\rm bb}  = 6 b_1 (D+4) A_{\text{1b}}+\frac{3}{2} b_2 b_1^2 (D+4)^2+6 \text{I}_{\text{bball}} \nn \\
&& ~~~~~~ -\frac{3 b_1^2 (d-3) (D+4) (D-d) J_{\text{M1a}}}{2 M^2}+3 b_2 b_1 (d-4) (D+4)
   J_{\text{m1a}}+12 b_2 b_1 (D+4) J_{\text{M1a}}+6 J_{\text{M3b}} \nn \\
&& ~~~~~~ -24 J_{\text{M3e}}+24
   J_{\text{M3f}} -6 (d-3) J_{\text{m1a}} J_{\text{M2a}}-36 (d-3) J_{\text{M1a}} J_{\text{M2a}} \nn \\
&& ~~~~~~ -\frac{9 (d-3) J_{\text{m1a}} J_{\text{M1a}}^2}{M^2}-\frac{3 (d-3) J_{\text{m1a}}^2
   J_{\text{M1a}}}{4 M^2}-\frac{27 (d-3) J_{\text{M1a}}^3}{M^2}\nn\\
%%%%%%%%%%%%%%%%%%%%%%%%%%%%%%%%%%%%%%%%%%%%%
&& B_{\rm sc}^{\rm bf} = -48 f_1 A_{\text{1b}}-12 b_1 (D+4) A_{\text{1f}}-24 b_1 b_2 (D+4) f_1-24
   \text{I}_{\text{fball}} \nn \\
&& ~~~~~~ -48 \text{I}_{\text{2c}} J_{\text{m1a}}+\frac{12 b_1 (d-3) f_1 (D-d) J_{\text{M1a}}}{M^2}-24
   b_2 (d-4) f_1 J_{\text{m1a}}-96 b_2 f_1 J_{\text{M1a}} \nn \\
&& B_{\rm sc}^{\rm ff} = 
96 f_1 A_{\text{1f}}+96 b_2 f_1^2+24 \text{I}_{\text{ffball}} \,.\nn 
\eea
\section{Diagrams}

We use conventional notation: spiral lines for gluons, dotted lines for ghosts, dashed lines for scalars, and solid lines for fermions. 

\newpage

\par
\newpage

\par

\newpage

%-----------------------------------------------------------------------
\section*{Acknowledgments}
%-----------------------------------------------------------------------

MEC thanks Jens Andersen for helpful conversations. 
This work was partially supported by the Natural Sciences and Engineering Research Council of Canada under grant SAPIN-2023-00023.

%\newpage

\end{document}